\documentclass[prd,eqsecnum,amsfonts,floatfix,nofootinbib,superscriptaddress,twocolumn]{revtex4}
\usepackage[pdftex]{graphicx}
\usepackage{amsmath}
\usepackage{aeitensor}
\usepackage{multirow}
\usepackage{url}
\usepackage{color}
\usepackage{hyperref}
\newcommand{\aref}[1]{appendix~\ref{#1}}
\newcommand{\sref}[1]{section~\ref{#1}}
\newcommand{\fref}[1]{figure~\ref{#1}}

\newcommand{\eref}[1]{\eqref{#1}}
\newcommand{\Aref}[1]{Appendix~\ref{#1}}
\newcommand{\Sref}[1]{Section~\ref{#1}}
\newcommand{\Fref}[1]{Figure~\ref{#1}}

\newcommand{\uhat}{\vec{u}}
\newcommand{\vhat}{\vec{v}}
\newcommand{\ihat}{\vec{\imath}}
\newcommand{\jhat}{\vec{\jmath}}
\newcommand{\khat}{\vec{k}}
\newcommand{\lhat}{\vec{\ell}}
\newcommand{\mhat}{\vec{m}}
\newcommand{\cosf}{\cos\phi_0}
\newcommand{\sinf}{\sin\phi_0}

\newcommand{\cost}{\cos 2\psi}
\newcommand{\sint}{\sin 2\psi}

\newcommand{\cosfpt}{\cos(\phi_0+2\psi)}
\newcommand{\sinfpt}{\sin(\phi_0+2\psi)}
\newcommand{\cosfmt}{\cos(\phi_0-2\psi)}
\newcommand{\sinfmt}{\sin(\phi_0-2\psi)}
\newcommand{\Rx}{\A^{\onedot}}
\newcommand{\Ry}{\A^{\twodot}}
\newcommand{\Lx}{\A^{\tredot}}
\newcommand{\Ly}{\A^{\fordot}}
\newcommand{\Rxp}{\A^{\widehat{1}}}
\newcommand{\Ryp}{\A^{\widehat{2}}}
\newcommand{\Left}{{\textsc{l}}}
\newcommand{\Right}{{\textsc{r}}}
\newcommand{\AR}{A_\Right}
\newcommand{\AL}{A_\Left}
\newcommand{\tR}{\phi_\Right}
\newcommand{\tL}{\phi_\Left}
\newcommand{\rR}{r_\Right}
\newcommand{\rL}{r_\Left}
\newcommand{\xR}{x_\Right}
\newcommand{\yR}{y_\Right}
\newcommand{\xL}{x_\Left}
\newcommand{\yL}{y_\Left}
\newcommand{\MLERx}{\MLE{\A}^{\onedot}}
\newcommand{\MLERy}{\MLE{\A}^{\twodot}}
\newcommand{\MLELx}{\MLE{\A}^{\tredot}}
\newcommand{\MLELy}{\MLE{\A}^{\fordot}}
\newcommand{\MLEAR}{\MLE{A}_\Right}
\newcommand{\MLEAL}{\MLE{A}_\Left}
\newcommand{\MLEtR}{\MLE{\phi}_\Right}
\newcommand{\MLEtL}{\MLE{\phi}_\Left}
\newcommand{\LambdaR}{\Lambda_\Right}
\newcommand{\LambdaL}{\Lambda_\Left}
\newcommand{\alphaR}{\alpha_\Right}
\newcommand{\alphaL}{\alpha_\Left}
\newcommand{\BR}{\B_\Right}
\newcommand{\BL}{\B_\Left}
\newcommand{\FsR}{\F_\Right}
\newcommand{\FsL}{\F_\Left}
\newcommand{\mc}[1]{\mathcal{#1}}
\newcommand{\Ap}{A_{+}}
\newcommand{\Ac}{A_{\times}}
\newcommand{\HA}{H}
\newcommand{\e}{\eta}
\newcommand{\coshe}{\cosh\e}
\newcommand{\sinhe}{\sinh\e}

\newcommand{\ep}{\tens{e}_{\!+}}
\newcommand{\ec}{\tens{e}_{\!\times}}
\newcommand{\epsp}{\tens{\varepsilon}_{\!+}}
\newcommand{\epsc}{\tens{\varepsilon}_{\!\times}}
\newcommand{\epsr}{\tens{\varepsilon}_{\!\Right}}
\newcommand{\epsl}{\tens{\varepsilon}_{\!\Left}}
\newcommand{\cosi}{\chi}
\newcommand{\A}{\mc{A}}
\newcommand{\MLE}[1]{\widehat{#1}}
\newcommand{\mbe}{\mathbin{=}}
\newcommand{\M}{\mc{M}}
\newcommand{\circidx}[1]{\breve{#1}}
\newcommand{\mudot}{\circidx{\mu}}
\newcommand{\nudot}{\circidx{\nu}}
\newcommand{\lamdot}{\circidx{\sigma}}
\newcommand{\lam}{\sigma}
\newcommand{\onedot}{\circidx{1}}
\newcommand{\twodot}{\circidx{2}}
\newcommand{\tredot}{\circidx{3}}
\newcommand{\fordot}{\circidx{4}}
\newcommand{\muhat}{\widehat{\mu}}
\newcommand{\nuhat}{\widehat{\nu}}
\newcommand{\F}{\mc{F}}
\newcommand{\B}{\mc{B}}
\newcommand{\Hf}{\mc{H}_s}
\newcommand{\Hi}{\mc{H}_s}
\newcommand{\abs}[1]{\left\lvert#1\right\rvert}

\newcommand{\tens}[1]{\aeitensor{#1}}

\newcommand{\cft}[1]{\widetilde{#1}}
\newcommand{\tssb}{\tau}

\DeclareMathOperator{\Real}{Re}
\DeclareMathOperator{\Imag}{Im}
\newcommand{\pdf}{\text{pdf}}
\newcommand{\Tsft}{T_{\text{sft}}}
\newcommand{\hdet}{h_\text{det}}

\newcommand{\coord}{coordinate}
\newcommand{\Coord}{Coordinate}

\newcommand{\coeff}{coefficient}

\newcommand{\iSFT}{l}
\newcommand{\sumXiSFT}{\sum_{X\iSFT}}
\newcommand{\scalar}[2]{\left(#1|#2\right)}
\newcommand{\detV}[1]{{#1}}
\newcommand{\detVm}[1]{{#1}}

\newcommand{\dcc}{LIGO-P1300105-v5}
\newcommand{\aei}{AEI-2013-250}

\def\commitDATE{ Thu Jan 23 06:36:08 2014 -0500}

\begin{document}
\title{New {\Coord}s for the Amplitude Parameter Space\\
  of Continuous Gravitational Waves}
\author{John T.\ Whelan}
\email{john.whelan@astro.rit.edu}
\affiliation{Center for Computational Relativity and Gravitation
  and School of Mathematical Sciences, Rochester Institute of Technology,
  85 Lomb Memorial Drive, Rochester, NY 14623, USA}
\author{Reinhard Prix}
\email{reinhard.prix@aei.mpg.de}
\affiliation{Max-Planck-Institut f\"{u}r Gravitationsphysik
  (Albert-Einstein-Institut), D-30167 Hannover, Germany}
\author{Curt J.\ Cutler}
\email{Curt.J.Cutler@jpl.nasa.gov}
\affiliation{Jet Propulsion Laboratory, M/S 169-327, 4800 Oak Grove Drive,
  Pasadena, CA 91109, USA}
\author{Joshua L.\ Willis}
\email{josh.willis@acu.edu}
\affiliation{Max-Planck-Institut f\"{u}r Gravitationsphysik
  (Albert-Einstein-Institut), D-30167 Hannover, Germany}
\affiliation{Department of Engineering \& Physics, Abilene Christian University, ACU
  Box 27963 Abilene, TX 79699, USA}
\date{\commitDATE%
}
\begin{abstract}
  The parameter space for continuous gravitational waves can be
  divided into amplitude parameters (signal amplitude, inclination and
  polarization angles describing the orientation of the source, and an
  initial phase) and phase-evolution parameters (signal frequency and
  frequency derivatives, and parameters such as sky position which
  determine the Doppler modulation of the signal).  The division is
  useful in part because of the existence of a set of functions known
  as the Jaranowski-Kr\'{o}lak-Schutz (JKS) {\coord}s, which are a set
  of four {\coord}s on the amplitude parameter space such that the
  gravitational-wave signal can be written as a linear combination of
  four template waveforms (which depend on the phase-evolution
  parameters) with the JKS {\coord}s as {\coeff}s.  We define a new
  set of {\coord}s on the amplitude parameter space, with the same
  properties, which can be more closely connected to the physical
  amplitude parameters.  These naturally divide into two pairs of
  Cartesian-like {\coord}s on two-dimensional subspaces, one
  corresponding to left- and the other to right-circular polarization.
  We thus refer to these as CPF (circular polarization
  factored) {\coord}s.  The corresponding two sets of polar {\coord}s
  (known as CPF-polar) can be related in a simple way to the physical
  parameters.  A further {\coord} transformation can be made, within
  each subspace, between CPF and so-called root-radius {\coord}s,
  whose radial {\coord} is the fourth root of the radial {\coord} in
  CPF-polar {\coord}s.  We illustrate some simplifying applications
  for these various {\coord} systems, such as: a calculation of the
  Jacobian for the transformation between JKS or CPF {\coord}s and
  the physical amplitude parameters (amplitude, inclination,
  polarization and initial phase); a demonstration that the Jacobian
  between root-radius {\coord}s and the physical parameters is a
  constant; an illustration of the signal {\coord} singularities
  associated with left- and right-circular polarization, which
  correspond to the origins of the two two-dimensional subspaces; and
  an elucidation of the form of the log-likelihood ratio between
  hypotheses of Gaussian noise with and without a continuous
  gravitational-wave signal.  These are used to illustrate some of the prospects
  for approximate evaluation of a Bayesian detection statistic defined
  by marginalization over the physical parameter space.  Additionally,
  in the presence of simplifying assumptions about the observing
  geometry, we are able, using CPF-polar {\coord}s, to explicitly
  evaluate the integral for the Bayesian detection statistic, and
  compare it to the approximate results.
\end{abstract}
\preprint{\dcc}
\maketitle

\section{Overview}

The gravitational-wave (GW) signal emitted from a nearly periodic,
non-precessing system, such as a rotating neutron star or a
slowly-evolving binary of compact objects, can be described by a
number of system parameters, such as the sky position (e.g., as
described by right ascension and declination) of the source, the
instantaneous frequency of the signal as a function of time, the
orientation of the equatorial/orbital plane, the distance to the
source, etc.  Four of these parameters (a combination of distance,
moments of inertia and frequency known as the signal amplitude $h_0$,
an initial phase $\phi_0$, the inclination $\iota$ of the system
angular momentum to the line of sight, and a polarization angle $\psi$
which describes the orientation of the equatorial/orbital plane) are generally
known as amplitude parameters (or sometimes extrinsic parameters).
Jaranowski et al \cite{jks98:_data} showed that the
GW signal can be written as a linear combination of four template
waveforms, with {\coeff}s given by four functions of the amplitude
parameters $\{h_0,\cosi\mbe\cos\iota,\psi,\phi_0\}$ and the form of the template
waveforms depending on the remaining parameters, known variously as
phase parameters, Doppler parameters, or intrinsic parameters.  (We
refer to them as phase-evolution parameters.)  The log-likelihood
ratio between models including Gaussian noise with and without a
continuous GW signal is then quadratic in these four functions, known
as the Jaranowski-Kr\'{o}lak-Schutz (JKS) {\coord}s on amplitude
parameter space.  This allows the likelihood function to be maximized
analytically over these parameters, which forms the basis of the
$\F$-statistic method \cite{jks98:_data} to search for continuous
gravitational waves. Prix and Krishnan \cite{Prix09:_Bstat} propose an
alternative, Bayesian-inspired detection statistic, in which the
likelihood ratio is marginalized over the amplitude parameters to
generate a Bayes factor to compare the signal and noise hypotheses.
The specific form of this statistic, known as the $\B$-statistic,
depends on the prior probability distribution for the amplitude
parameters.  Taking a prior distribution uniform in the JKS {\coord}s
would produce a statistic equivalent to the $\F$-statistic.
However, a physically realistic prior distribution should be isotropic
in the orientation of the equatorial/orbital plane of the emitting
system, i.e., uniform in both $\cosi=\cos\iota$ and $\psi$.  Thus the
relationship between the JKS and physical {\coord}s is important for
evaluating the $\B$-statistic, either in JKS {\coord}s, where the
likelihood ratio is a Gaussian but the prior probability distribution
is more complicated, or in physical {\coord}s, where the prior is
simple by the likelihood is more complicated.  This paper proposes
several new sets of {\coord}s on amplitude parameter space which
elucidate this relationship, and the relationship between physical
amplitude parameters and the GW signal.

The paper is organized as follows: In \sref{s:signal} we write the
explicit signal model for continuous GWs, in terms both of the tensor
GW propagating from the source to the solar system and of the signal
received in each detector, indicating the dependence on the amplitude
and phase-evolution parameters.

In \sref{s:coords} we describe three types of {\coord}s on the
amplitude parameter space: the physical {\coord}s
$\{h_0,\iota,\psi,\phi_0\}$ related to the emitting system, the JKS
{\coord}s $\{\mc{A}^1,\mc{A}^2,\mc{A}^3,\mc{A}^4\}$ in which the
signal is linear, and a new set of {\coord}s
$\{\Rx,\Ry,\Lx,\Ly\}$ which also have this
property, but are more simply related to the physical {\coord}s.
Because the {\coord} pairs $\{\Rx,\Ry\}$ and
$\{\Lx,\Ly\}$ span the space of right- and left-handed
circular polarization, respectively, we refer to
$\{\Rx,\Ry,\Lx,\Ly\}$ as CPF (circular
polarization factored) {\coord}s.  Considering the
combinations $\{\Rx,\Ry\}$ and $\{\Lx,\Ly\}$ as
Cartesian {\coord}s on their respective two-dimensional subspaces, we
define the corresponding polar {\coord}s $\{\AR,\tR\}$ and
$\{\AL,\tL\}$--known as CPF-polar {\coord}s--which have the practical
advantage that $\{\AR,\AL\}$ are functions of only $\{h_0,\iota\}$ and
$\{\tR,\tL\}$ are functions of only $\{\psi,\phi_0\}$.  A final useful
{\coord} transformation is to so-called root-radius {\coord}s which
use the same angles $\{\tR,\tL\}$ but define radial {\coord}s
$\rR=\AR^{1/4}$ and $\rL=\AL^{1/4}$.  The root-radius {\coord}s have
corresponding Cartesian counterparts defined from the polar pairs in the
usual way.

In \sref{s:apps} we illustrate several simple applications of these
new {\coord}s: \Sref{s:apps-jacobian} contains a simple analytic
calculation of the Jacobian of the transformation between JKS and
physical {\coord}s, previously calculated in \cite{Prix09:_Bstat}
using the symbolic manipulation program \textsc{maxima}\cite{maxima}.
We also illustrate the Jacobians for conversions between various sets
of {\coord}s and show that the Jacobian between physical and
root-radius {\coord}s is a constant.  In \sref{s:apps-circpol} we
consider the nature of the {\coord} singularities associated with
right and left circular polarization, which correspond to $\AR=0$ and
$\AL=0$, respectively.

\Sref{s:Bstat} contains several illustrations of how the new {\coord}s
can be applied to computation of the $\B$-statistic, by writing,
in \sref{s:Bstat-loglike}, the log-likelihood ratio explicitly in the
new {\coord}s.
In \sref{s:Bstat-sigcoord} we illustrate the problem with
an obvious technique for approximate calculation of the
$\B$-statistic integral in JKS or CPF {\coord}s, i.e., Taylor
expanding the logarithm of the Jacobian appearing in the prior
probability density function (pdf)
about the maximum-likelihood point.  The problem is that the resulting
Gaussian expression does not always have a maximum at the expected point;
if the maximum likelihood
signal parameters are too close to circular polarization, the
integrand has a saddle point at the point of interest, not a maximum.
In \sref{s:Bstat-rootrad} we show that an approximate Gaussian
integration \emph{can} be performed in root-radius {\coord}s, which
gives a simple relationship between the $\B$-statistic and the
$\F$-statistic, which should be valid when both the left- and
right-circular polarization amplitudes are large compared to the scale
set by the detector sensitivity and observing time.
In \sref{s:Bstat-phys}, the expression for the log-likelihood in the
new {\coord}s of this paper is used to simplify evaluation of the
$\B$-statistic integral as an integral over the physical parameters by
making clear the dependence on the $h_0$ and $\phi_0$ parameters, the
integrals over which can be performed analytically.

In \sref{s:explicit} we consider the special case where the averaged
amplitude-modulation {\coeff}s have a simple form which causes the
likelihood ratio to factor into pieces related to the two
circular-polarization subspaces.  In that case, the $\B$-statistic
integral can be evaluated explicitly in CPF-polar {\coord}s.  We
compare the exact solution to the various approximations considered in
\sref{s:Bstat}.

\Aref{app:metric} spells out the calculation of the log-likelihood
ratio in CPF {\coord}s, in particular the ``metric'' made up of the
{\coeff}s in the quadratic terms.  \Aref{app:hyp} contains another
related {\coord} system, which also have a constant Jacobian factor
relating them to the physical parameters, but whose practical
applications remain to be found.

\section[Signal Model]{Signal Model for Continuous Gravitational
  Waves}
\label{s:signal}

The tensor GW signal from a nearly periodic source can be written as
\begin{equation}
  \label{e:signal}
  \tens{h}(\tssb) = h_+(\tssb)\, \ep + h_\times(\tssb) \, \ec
\ ,
  \ ,
\end{equation}
with
\begin{subequations}
\begin{align}
  \label{eq:signal2}
  h_+(\tssb) &\equiv \Ap \cos [\phi(\tssb)+\phi_0]
\ ,\\
  h_\times(\tssb) &\equiv \Ac \sin [\phi(\tssb)+\phi_0] 
\ ,
\end{align}
\end{subequations}
where $\tssb$ is the time of arrival of the signal at the solar system
barycenter (SSB), $\ep$ and $\ec$ are polarization
basis tensors, $\Ap$ and $\Ac$ are the amplitudes of the corresponding
polarizations, $\phi(\tssb)$ describes the phase evolution of the
signal, and $\phi_0$ is the phase at the reference time $\tssb=0$.

\begin{figure}
  \begin{center}
    \includegraphics[width=\columnwidth]{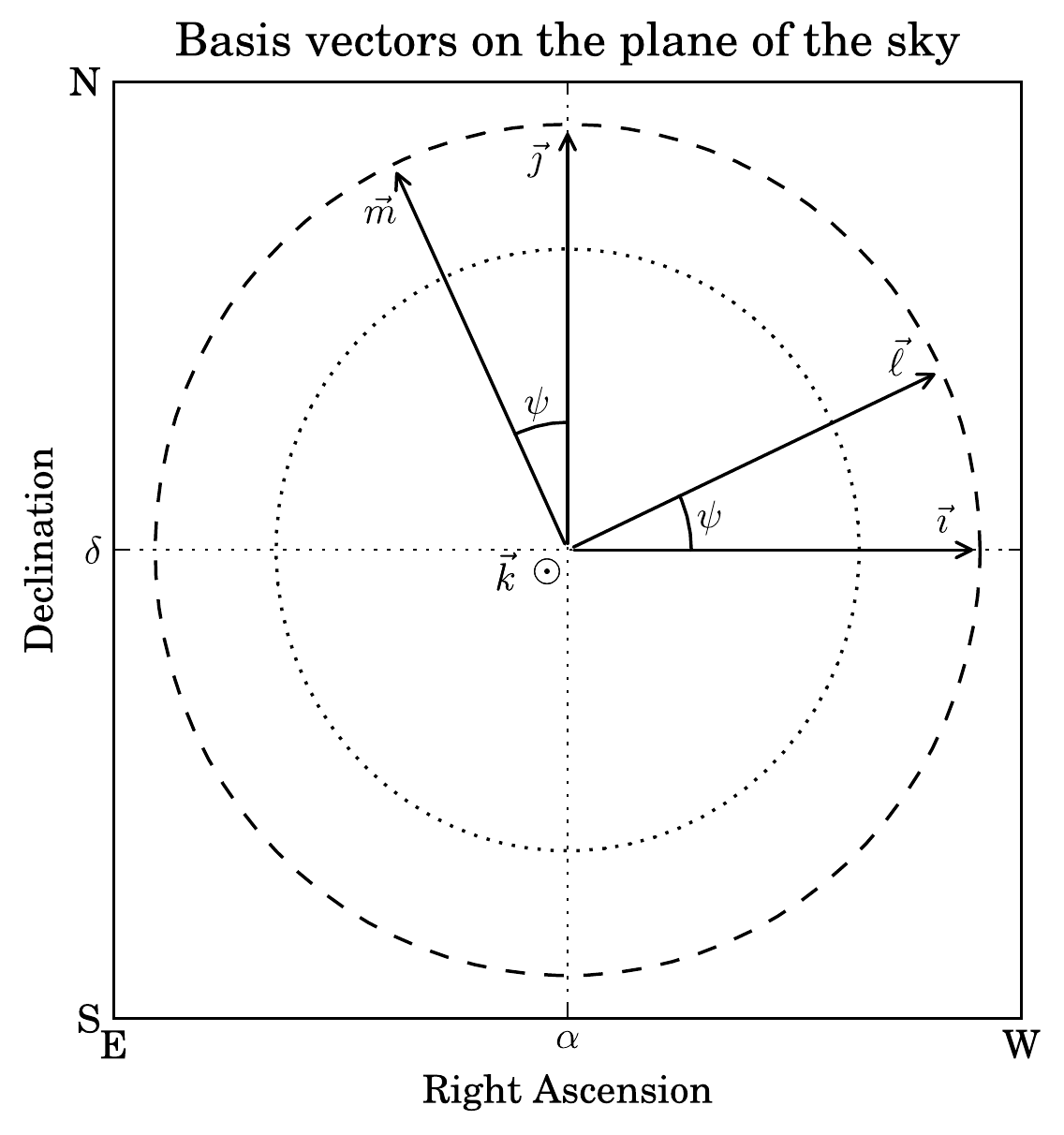}
  \end{center}
  \caption{Basis vectors used to define the basis tensors, from the
    perspective of the observer looking at the source.  The
    propagation unit vector $\khat$ is pointing out of the page.  The
    unit vector $\ihat$ lies in the equatorial plane perpendicular to
    the line of sight, pointing ``West on the sky'' in the direction
    of decreasing right ascension $\alpha$, and the unit vector
    $\jhat$ points ``North on the sky'' in the direction of increasing
    declination $\delta$.  They are used to construct basis tensors
    $\epsp = \ihat\otimes\ihat - \jhat\otimes\jhat$ and
    $\epsc = \ihat\otimes\jhat +
    \jhat\otimes\ihat$.  The source's preferred polarization basis
    $\ep = \lhat\otimes\lhat - \mhat\otimes\mhat$ and
    $\ec = \lhat\otimes\mhat + \mhat\otimes\lhat$ is
    defined by constructing a right-handed orthonormal basis
    $\{\lhat,\mhat,\khat\}$ such that either $\lhat$ or $\mhat$ to lie
    parallel or antiparallel to the projection onto the sky of the
    neutron star spin; the other unit vector then lies in the
    neutron star's equatorial plane.  The polarization
    angle $\psi$ is measured counterclockwise from $\ihat$ to $\lhat$.
    By choosing $\lhat$ or $\mhat$ to lie parallel or antiparallel to
    the projected angular momentum, $\psi$ can be arranged to lie in
    the interval $(-\pi/4,\pi/4]$.}
  \label{f:bases}
\end{figure}

If we denote the unit vector from the source to the SSB as $\khat$,
the polarization basis tensors can be constructed from unit vectors
which form a right-handed orthonormal basis $\{\lhat,\mhat,\khat\}$:
\begin{subequations}
  \label{e:basistensors}
  \begin{align}
    \ep &= \lhat\otimes\lhat - \mhat\otimes\mhat
    \\
    \ec &= \lhat\otimes\mhat + \mhat\otimes\lhat
    \ .
  \end{align}
\end{subequations}
Typical sources for GWs described by \eref{e:signal} are spinning
deformed neutron stars and slowly evolving compact binary systems.
For concreteness, we will refer to the former, but the signal geometry
is the same, with the equatorial plane of the spinning neutron star
replaced by the the orbital plane of the binary.  A polarization
basis which produces the signal \eref{e:signal}, in which the $+$ and
$\times$ components are a quarter-cycle out of phase, is obtained by
choosing either $\lhat$ or $\mhat$ to lie in the equatorial plane of
the neutron star.  For a given source sky position (which can be
specified by right ascension $\alpha$ and declination $\delta$, and
defines the propagation direction $\khat$ from the source to the SSB),
we need an additional polarization angle $\psi$ to specify the
orientation of the basis vectors $\{\lhat,\mhat\}$ used to construct
the polarization basis tensors $\{\ep,\ec\}$, as
illustrated in \fref{f:bases}.
The angle $\psi$ is measured counter-clockwise from a reference
direction $\ihat$ to $\lhat$.  The freedom to choose $\lhat$ or
$\mhat$ pointing in either direction within the orbital plane allows
us to restrict $\psi$ to a 90-degree interval such as
$(-\pi/4,\pi/4]$.  The reference direction $\ihat$ is defined to lie
in the Earth's equatorial plane, perpendicular to the line of sight,
pointing in the local ``West on the sky'' direction of decreasing
right ascension.  Together with a unit vector $\jhat$ pointing ``North
on the sky'' (perpendicular to the line of sight, in the direction of
increasing declination), it forms a right-handed orthonormal basis
$\{\ihat,\jhat,\khat\}$.  We can use this basis to form an alternate
set of basis tensors
\begin{subequations}
  \label{e:earthtensors}
  \begin{align}
    \epsp &= \ihat\otimes\ihat - \jhat\otimes\jhat
    \\
    \epsc &= \ihat\otimes\jhat + \jhat\otimes\ihat
  \end{align}
\end{subequations}
which are determined entirely by the sky position $\{\alpha,\delta\}$.
In terms of this alternate polarization basis, the preferred basis can
be written as
\begin{subequations}
  \label{e:tensorrot}
  \begin{alignat}{3}
    \ep\, &= &\epsp&\,\cos2\psi\,
    &+&\,\epsc\,\sin2\psi
    \\
    \ec\, &= -\,&\epsp&\,\sin2\psi\,
    &+&\,\epsc\,\cos2\psi
    \ .
  \end{alignat}
\end{subequations}

For GWs generated by a non-precessing system with
nearly periodically varying quadrupole moments (e.g., a triaxial
neutron star spinning about a principal axis), the amplitudes of the
two polarizations are given by
\begin{subequations}
  \begin{align}
    \Ap &= \frac{h_0}{2} (1+\cos^2\iota)
    \\
    \Ac &= h_0 \cos\iota
    \ ,
  \end{align}
\end{subequations}
where $\iota$ is the angle between the line of sight and the neutron
star's rotation axis, and
\begin{equation}
  h_0 = \frac{4G}{c^4}\frac{\abs{I_{xx}-I_{yy}}\Omega^2}{d}
\end{equation}
is the amplitude in terms of the equatorial quadrupole moments
$\{I_{xx},I_{yy}\}$, the rotation frequency $\Omega$, and the distance
$d$ to the source.

Finally, the phase evolution $\phi(\tssb)$ at the SSB is typically
described in terms of parameters describing the neutron star rotation
and spindown, e.g.,
\begin{equation}
  \phi(\tssb) = 2\pi \left(f_0\tssb + f_1\frac{\tssb^2}{2} + \cdots\right)
  \ ,
\end{equation}
although it may be more complicated if, e.g., the spinning neutron star
is in a binary system which Doppler modulates the signal.

The parameters describing the signal are divided into two categories:
\begin{itemize}
\item \textit{Amplitude parameters} $\{h_0,\iota,\psi,\phi_0\}$, and
\item \textit{Phase-evolution parameters} such as the sky position
  $\{\alpha,\delta\}$, signal frequency and spindown parameters
  ${f_0,f_1,\ldots}$, and any orbital parameters for spinning neutron
  stars in binary systems.
\end{itemize}

Finally, the measured signal $h^X(t)$ at time $t$ by detector $X$ is
the response of the detector to the GW tensor
$\tens{h}^X(t)\equiv\tens{h}(\tssb^X(t))$.
The function $\tssb^X(t)$ denotes the SSB arrival time $\tssb$ of a
wavefront that reaches detector $X$ at time $t$, which accounts for
the sky-position $\{\alpha,\delta\}$ dependent Doppler modulation due
to detector motion.

If we consider a stretch of time that is short enough for the detector
arms to have approximately constant orientation, then we can most
easily write the general detector response in the frequency domain
(see for example \cite{Rubbo:2003ap,WhelanPrix08:_MLDC1B}) as
\begin{equation}
  \begin{split}
    \cft{h}^X(f) &= \cft{\tens{h}}{}^X(f):\tens{d}^X(f) \label{e:response}
    \\
    &= \cft{h}_+^X(f) \, F_+^X(f) + \cft{h}_\times^X(f)\, F_\times^X(f)
    \ ,
  \end{split}
\end{equation}
where $\cft{\;\;\;}$ denotes the Fourier-transform, and the antenna
pattern functions are defined as
\begin{subequations}
\begin{align}
  F_+^X(f) &\equiv \ep:\tens{d}^X(f)
\ ,\\
  F_\times^X(f) &\equiv \ec:\tens{d}^X(f)
\ ,
\end{align}
\end{subequations}
in terms of the (generally complex, and sky-position dependent)
detector tensor $\tens{d}^X(f)$.
Along the lines of \eref{e:tensorrot}, the dependence of
$\{F_+,F_\times\}$ upon the sky position and the polarization basis
can be separated as
\begin{subequations}
  \begin{alignat}{3}
    F_+(\alpha,\delta,\psi)\, &= &a(\alpha,\delta)&\,\cos2\psi\,
    &+&\,b(\alpha,\delta)\,\sin2\psi
\ ,
    \\
    F_\times(\alpha,\delta,\psi)\, &= -\,&a(\alpha,\delta)&\,\sin2\psi\,
    &+&\,b(\alpha,\delta)\,\cos2\psi
\ ,
  \end{alignat}
\end{subequations}
in terms of the (generally complex) amplitude modulation {\coeff}s
\begin{subequations}
  \begin{align}
  a^X(f) &\equiv \epsp:\tens{d}^X(f)
\ ,\\
  b^X(f) &\equiv \epsc:\tens{d}^X(f)
\ ,
  \end{align}
\label{eq:def-a-b}
\end{subequations}
which are independent of the signal amplitude parameters.

In the case of ground-based detectors, one commonly uses the
long-wavelength limit approximation, as the interferometer arms are
typically much shorter than the wavelength $c/f_0$ of the GWs.
In this limit the detector-response tensor $\tens{d}(f)$ becomes
real-valued and independent of frequency (and sky-position), and can
be expressed as
\begin{equation}
  \label{e:dLWL}
  \tens{d}(f) \approx \tens{d}_{\mathrm{LWL}}\equiv \frac{1}{2}\left(\uhat\otimes\uhat + \vhat\otimes\vhat\right)
\ ,
\end{equation}
for interferometer arms along unit vectors $\uhat$ and $\vhat$.

\section[Amplitude {\Coord}s]{{\Coord}s on Amplitude Parameter Space}
\label{s:coords}

\subsection{Physical {\Coord}s}

The amplitude parameters most closely connected to the geometry of the
emitting system are $\{h_0,\iota,\psi,\phi_0\}$.  They form a set of {\coord}s
on the four-dimensional amplitude parameter space.  Any signal of the
form \eref{e:signal} can be described by parameters in the range
\begin{equation}
  0 \le h_0 < \infty
  \qquad\hbox{and}\qquad
  0 \le \iota \le \pi
\end{equation}
and
\begin{equation}
  \label{e:fprange}
  -\frac{\pi}{4} < \psi \le \frac{\pi}{4}
  \qquad\hbox{and}\qquad
  0 \le \phi_0 < 2\pi
\end{equation}
The range of angles can be understood by noting that if we make the
transformation $\psi\rightarrow\psi + \pi/2$, \eref{e:tensorrot}
implies that $\{\ep,\ec\}\rightarrow
\{-\ep,-\ec\}$, which means that the transformation
$\{\psi,\phi_0\}\rightarrow\{\psi + \pi/2,\phi_0+\pi\}$ leaves the
waveform \eref{e:signal} unchanged.

It is also convenient to define $\cosi=\cos\iota$, so that
\begin{equation}
  \Ap = \frac{h_0}{2} (1+\cosi^2)
  \qquad\hbox{and}\qquad
  \Ac = h_0 \cosi
\end{equation}
and consider the physical {\coord}s
$\{h_0,\cosi\mbe\cos\iota,\psi,\phi_0\}$ with parameter space ranges
\begin{equation}
  \label{e:hcrange}
  0 \le h_0 < \infty
  \qquad\hbox{and}\qquad
  -1 \le \cosi \le 1
\end{equation}
If the distribution of neutron star spins is isotropic, a physical
probability distribution on amplitude parameter space should be
uniform in $\cosi$ and $\psi$ as well as $\phi_0$, so that
\begin{equation}
  \label{e:physpdf}
  \pdf(h_0,\cosi,\psi,\phi_0|\Hi) = \frac{1}{2\pi^2}\pdf(h_0|\Hi)
  \quad \hbox{(isotropic prior)}
\end{equation}

Finally, note that the range on $h_0$ and $\cosi$ implies that
\begin{equation}
  0 \le \Ap < \infty
  \qquad\hbox{and}\qquad
  -\Ap \le \Ac \le \Ap
\end{equation}

If $\iota=\pi/2$, so that $\cosi=0$, the GW signal is \textit{linearly
  polarized}.  In this case, $\Ap=h_0/2$, $\Ac=0$, and the signal in
the preferred basis contains only the plus polarization state:
\begin{equation}
  \begin{split}
    \tens{h}(\tssb) &= \frac{h_0}{2}\cos(\phi(\tssb)+\phi_0) \ep
    \\
    &= \frac{h_0}{2}\cos(\phi(\tssb)+\phi_0)
    \left(
      \epsp\,\cos2\psi+\epsc\,\sin2\psi
    \right)
  \end{split}
\end{equation}

If $\iota=0$, so that $\cosi=1$, the GW signal is \textit{right
  circularly polarized}.  In this case, $\Ap=h_0=\Ac$ and the signal
is
\begin{equation}
  \begin{split}
    \tens{h}(\tssb)
    &= h_0
    \left[
      \cos(\phi(\tssb)+\phi_0) \ep
      + \sin(\phi(\tssb)+\phi_0) \ec
    \right]
    \\
    &=
    h_0
    \left[
      \cos(\phi(\tssb)+\phi_0-2\psi) \epsp
      + \sin(\phi(\tssb)+\phi_0-2\psi) \epsc
    \right]
  \end{split}
\end{equation}
We see that for right circular polarization there is a degeneracy of
the $\psi$ and $\phi_0$ {\coord}s, with the waveform depending only on
the combination $\phi_0-2\psi$.

If $\iota=\pi$, so that $\cosi=-1$, the GW signal is \textit{left
  circularly polarized}.  In this case, $\Ap=h_0=-\Ac$ and the signal
is
\begin{equation}
  \begin{split}
    \tens{h}(\tssb)
    &= h_0
    \left[
      \cos(\phi(\tssb)+\phi_0) \ep
      - \sin(\phi(\tssb)+\phi_0) \ec
    \right]
    \\
    &=
    h_0
    \left[
      \cos(\phi(\tssb)+\phi_0+2\psi) \epsp
      + \sin(\phi(\tssb)+\phi_0+2\psi) \epsc
    \right]
  \end{split}
\end{equation}
We see that for right circular polarization there is a degeneracy of
the $\psi$ and $\phi_0$ {\coord}s, with the waveform depending only on
the combination $\phi_0+2\psi$.

\subsection{JKS $\A^\mu$ {\Coord}s}
\label{s:JKScoords}

The basis of the $\F$-statistic maximum likelihood method
\cite{jks98:_data} is the observation that the GW signal
\eref{e:signal} is linear in the following four combinations of the
four amplitude parameters:
\begin{subequations}
  \label{e:Amus}
  \begin{align}
    \A^1 &= \Ap\cost\cosf - \Ac\sint\sinf
    \\
    \A^2 &= \Ap\sint\cosf + \Ac\cost\sinf
    \\
    \A^3 &= - \Ap\cost\sinf - \Ac\sint\cosf
    \\
    \A^4 &= - \Ap\sint\sinf + \Ac\cost\cosf
    \ .
  \end{align}
\end{subequations}
The GW tensor waveform \eref{e:signal} can be written as
\begin{equation}
  \tens{h}(\tssb; \A, \lambda) = \A^\mu \tens{h}_\mu(\tssb;\lambda)
  \ ,
\end{equation}
where
\begin{subequations}
  \label{e:tenswf}
  \begin{align}
    \tens{h}_1(\tssb) &= \epsp\,\cos\phi(\tssb) \\
    \tens{h}_2(\tssb) &= \epsc\,\cos\phi(\tssb) \\
    \tens{h}_3(\tssb) &= \epsp\,\sin\phi(\tssb) \\
    \tens{h}_4(\tssb) &= \epsc\,\sin\phi(\tssb)
  \end{align}
\end{subequations}
and we have introduced the Einstein summation convention that sums
such as $\sum_{\mu=1}^4$ are implied when indices are repeated.  As
illustrated in \cite{Prix09:_Bstat}, using the maximized likelihood as
a detection statistic is equivalent to using a marginalized
likelihood, with an unphysical prior:
\begin{equation}
  \label{e:Fstatprior}
  \pdf(\A^1,\A^2,\A^3,\A^4|\Hf) = \text{constant}
  \qquad \hbox{($\F$-stat prior)}
\end{equation}
To convert the $\F$-statistic prior into physical {\coord}s, or to
convert a physical isotropic prior of the form \eref{e:physpdf} into
$\{\A^\mu\}$ {\coord}s requires the Jacobian for the transformation
between $\{h_0,\cosi,\psi,\phi_0\}$ and $\{\A^\mu\}$.  This was
reported in \cite{Prix09:_Bstat} as
\begin{equation}
  \label{e:dAdphys}
  d\A^1\,d\A^2\,d\A^3\,d\A^4 = \frac{h_0^3}{4}
  \left(1-\cosi^2\right)^3
  \,dh_0\,d\cosi\,d\psi\,d\phi_0
  \ ,
\end{equation}
a derivation of which we present in \sref{s:apps-jacobian}.
This means that, for example,
\begin{equation}
  \pdf(\A^1,\A^2,\A^3,\A^4|\Hi)
  = \frac{4\,\pdf(h_0,\cosi,\psi,\phi_0|\Hi)}
  {h_0^3\left(1-\cosi^2\right)^3}
  \ .
\end{equation}

\subsection{New {\Coord}s}

\subsubsection{CPF (Circular Polarization Factored) {\Coord}s}

We now introduce an alternate set of {\coord}s $\{\A^{\mudot}\}$ of
the form
\begin{subequations}
  \label{e:PQ}
  \begin{align}
    \Rx &\equiv \frac{\A^1 + \A^4}{2}
    \\
    \Ry &\equiv \frac{\A^2 - \A^3}{2}
    \\
    \Lx &\equiv \frac{\A^1 - \A^4}{2}
    \\
    \Ly &\equiv \frac{-\A^2 - \A^3}{2}
  \end{align}
\end{subequations}
The GW signal is also linear in these {\coord}s, with the form
(again using the Einstein summation convention)
\begin{equation}
  \label{e:tenswavedot}
  \tens{h}(\tssb;\A,\lambda) = \A^{\mudot} \,\tens{h}_{\mudot}(\tssb;\lambda)
  \ ,
\end{equation}
where
\begin{subequations}
  \label{e:dottenswf}
  \begin{align}
    \tens{h}_{\onedot} &= \tens{h}_1 + \tens{h}_4\\
    \tens{h}_{\twodot} &= \tens{h}_2 - \tens{h}_3 \\
    \tens{h}_{\tredot} &= \tens{h}_1 - \tens{h}_4 \\
    \tens{h}_{\fordot} &= - \tens{h}_2 - \tens{h}_3
  \end{align}
\end{subequations}
so these new {\coord}s can be used in an $\F$-statistic
construction in just the same way as the original JKS $\{\A^\mu\}$
{\coord}s.

The basis waveforms \eref{e:dottenswf} take on a simple form
if we define left- and right-circular polarization basis tensors as
\begin{equation}
  \label{e:circbasis}
  \epsr = \epsp + i\epsc
  \qquad\hbox{and}\qquad
  \epsl = \epsp - i\epsc
\end{equation}
then
\begin{subequations}
  \label{e:circtenswf}
  \begin{align}
    \tens{h}_{\onedot}(\tssb) &= \Real \left(\epsr\,e^{-i\phi(\tssb)}\right)
    \\
    \tens{h}_{\twodot}(\tssb) &= \Imag \left(\epsr\,e^{-i\phi(\tssb)}\right)
    \\
    \tens{h}_{\tredot}(\tssb) &= \Real \left(\epsl\,e^{-i\phi(\tssb)}\right)
    \\
    \tens{h}_{\fordot}(\tssb) &= \Imag \left(\epsl\,e^{-i\phi(\tssb)}\right)
  \end{align}
\end{subequations}
So we see that $\{\Rx,\Ry\}$ and $\{\Lx,\Ly\}$ are
amplitudes of the right- and left-circular polarized parts of the GW
signal, respectively, just as the JKS {\coord}s $\{\A^1,\A^3\}$ and
$\{\A^2,\A^4\}$ are amplitudes of the plus- and cross-polarized parts
of the GW signal in a particular polarization basis.  We thus refer to
$\{\A^{\mudot}\}
$ as circular polarization factored (CPF) {\coord}s.

The CPF {\coord}s are more closely connected to the physical
amplitude parameters than are the JKS {\coord}s.  In particular
\begin{subequations}
  \label{e:CPFphys}
  \begin{align}
    \Rx &= h_0\left(\frac{1+\cosi}{2}\right)^2\cosfpt
    \\
    \Ry &= h_0\left(\frac{1+\cosi}{2}\right)^2\sinfpt
    \\
    \Lx &= h_0\left(\frac{1-\cosi}{2}\right)^2\cosfmt
    \\
    \Ly &= h_0\left(\frac{1-\cosi}{2}\right)^2\sinfmt
    \ .
  \end{align}
\end{subequations}
Note that this decomposition can be written in terms of spin-weighted
spherical
harmonics\cite{GelfandMinlosShapiro,Newman:1966ub,Goldberg:1966uu,Thorne:1980ru}
as
\begin{subequations}
  \begin{align}
    \Rx + i \Ry = \AR e^{i\tR}
    &= \sqrt{\frac{4\pi}{5}} h_0 e^{i\phi_0} {}_{-2}Y_{2,+2}(\iota, \psi)
    \\
    \Lx + i \Ly = \AL e^{i\tL}
    &= \sqrt{\frac{4\pi}{5}} h_0 e^{i\phi_0} {}_{-2}Y_{2,-2}(\iota, \psi)
  \end{align}
\end{subequations}

\subsubsection{CPF-polar {\Coord}s}

\begin{figure*}
  \begin{center}
    \includegraphics[width=0.475\textwidth]{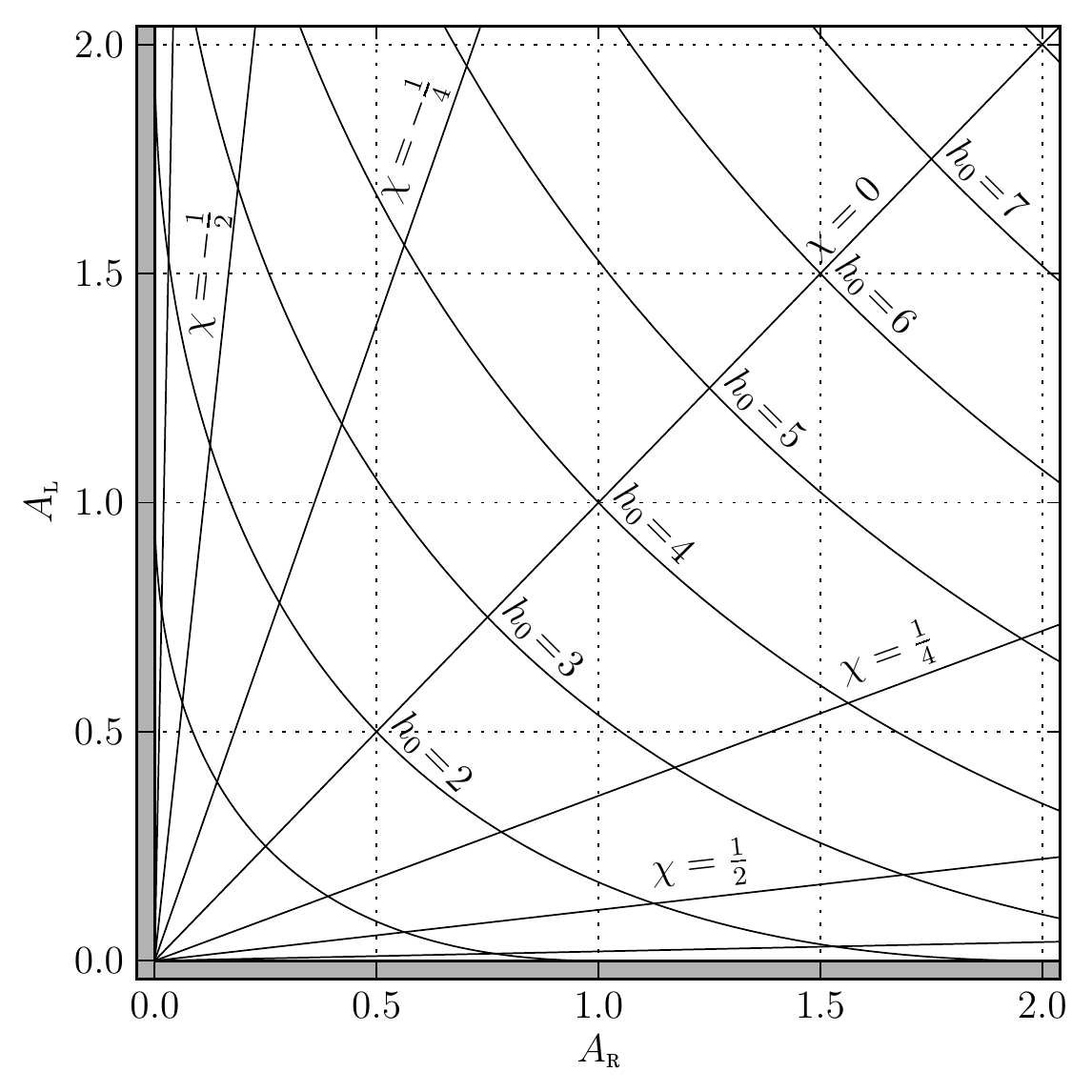}
    \includegraphics[width=0.475\textwidth]{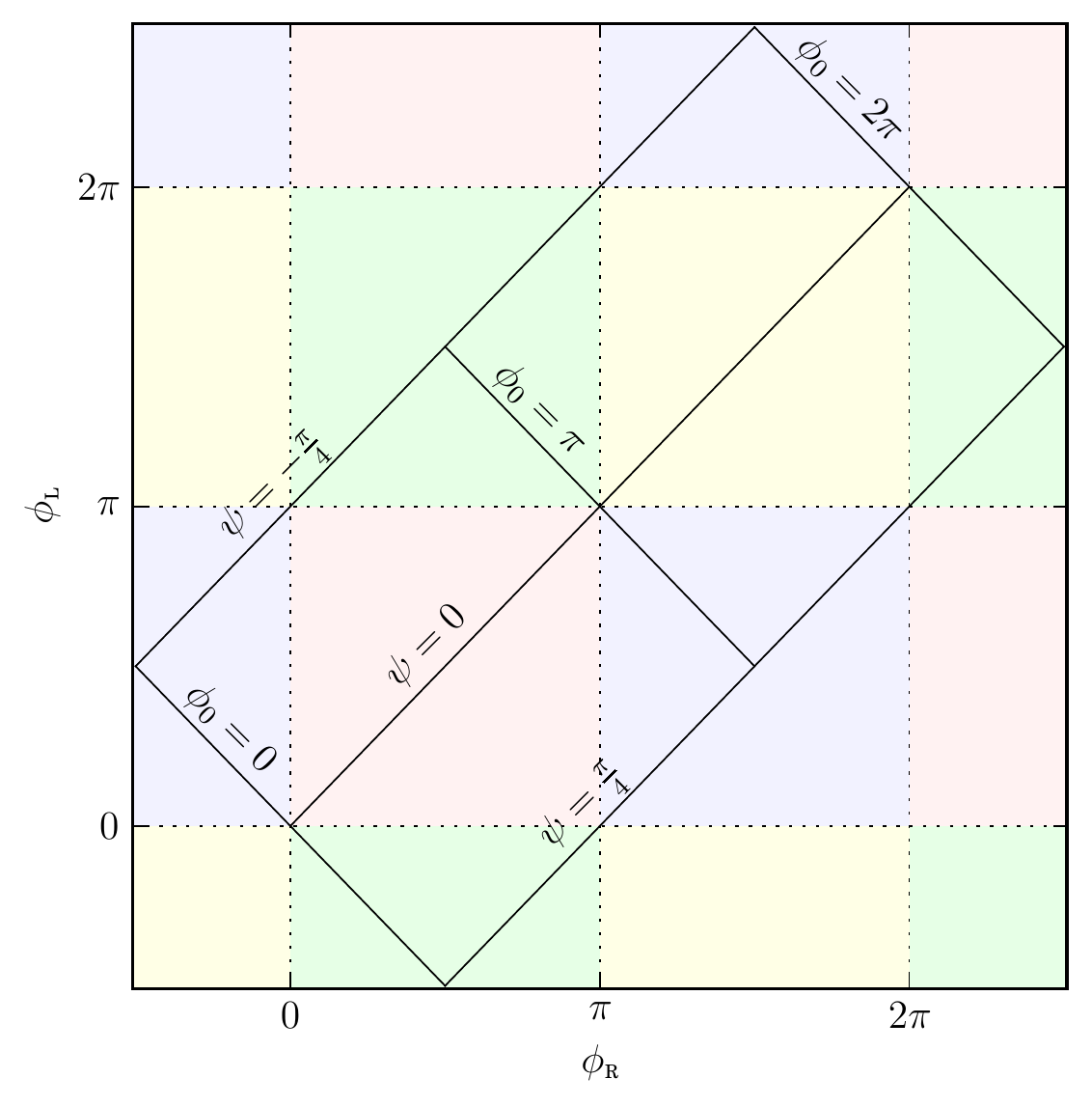}
  \end{center}
  \caption{Correspondence between $\{\AR,\tR\}$ and $\{\AL,\tL\}$, which
    act as polar {\coord}s for $\{\Rx,\Ry\}$ and
    $\{\Lx,\Ly\}$, respectively, and the physical amplitude
    parameters $\{h_0,\cosi\mbe\cos\iota,\psi,\phi_0\}$.  At left are
    lines of constant $h_0\in[0,\infty)$ and $\cosi\in[-1,1]$, drawn
    in first quadrant of the $\{\AR,\AL\}$ plane.  (The grey shaded
    region, where $\AR<0$ and/or $\AL<0$, represents unphysical {\coord}
    values.)  The positive $\AR$ axis, where $\AL=0$, corresponds to
    $\cosi=1$, where the GW signal is right circularly polarized.  The
    positive $\AL$ axis, where $\AR=0$, corresponds to $\cosi=-1$, where
    the GW signal is left circularly polarized.  At right, the
    principal region of polarization $\psi\in(-\pi/4,\pi/4]$ and phase
    $\phi_0\in[0,2\pi)$ is shown in the $\{\tR,\tL\}$ plane; $\tR$ and
    $\tL$ are each periodically identified, with period $2\pi$.  Note
    that since the transformation $\{\psi,\phi_0\}\rightarrow\{\psi +
    \pi/2,\phi_0+\pi\}$ leaves the waveform unchanged, the edge
    $\psi=-\pi/4$, $\phi_0\in[0,\pi)$ is actually identified with
    $\psi=\pi/4$, $\phi_0\in[\pi,2\pi)$, while $\psi=-\pi/4$,
    $\phi_0\in[\pi,2\pi)$ is identified with $\psi=\pi/4$,
    $\phi_0\in[0,\pi)$.  These periodic identifications show that the
    principal $\{\psi,\phi_0\}$ region is equivalent to the region
    $\tR\in[0,2\pi)$, $\tL\in[0,2\pi)$.}
  \label{f:grids}
\end{figure*}

The connection \eref{e:CPFphys} becomes even simpler if we introduce
polar {\coord}s on each of the two-dimensional subspaces:
\begin{subequations}
  \label{e:PQpolar}
  \begin{align}
    \Rx = \AR\cos\tR
    \qquad&\hbox{and}\qquad
    \Ry = \AR\sin\tR
    \\
    \Lx = \AL\cos\tL
    \qquad&\hbox{and}\qquad
    \Ly = \AL\sin\tL
    \ ;
  \end{align}
\end{subequations}
These {\coord}s, which we call CPF-polar {\coord}s, can be written
\begin{subequations}
  \label{e:pqphys}
  \begin{align}
    \AR = \frac{\Ap + \Ac}{2} = h_0\left(\frac{1+\cosi}{2}\right)^2
    \quad&\hbox{and}\quad
    \tR = \phi_0+2\psi
    \ ;
    \\
    \AL = \frac{\Ap - \Ac}{2} = h_0\left(\frac{1-\cosi}{2}\right)^2
    \quad&\hbox{and}\quad
    \tL = \phi_0-2\psi
    \ .
  \end{align}
\end{subequations}
We can see that \eref{e:hcrange} is equivalent to
\begin{equation}
  0 \le \AR < \infty
  \qquad\hbox{and}\qquad
  0 \le \AL < \infty
  \ ,
\end{equation}
while \eref{e:fprange} is equivalent, taking into account the
periodicity of the angles, to
\begin{equation}
  0 \le \tR < 2\pi
  \qquad\hbox{and}\qquad
  0 \le \tL < 2\pi
  \ ,
\end{equation}
which are just the ranges associated with $\{\AR,\tR\}$ and $\{\AL,\tL\}$
being polar {\coord}s.  The mapping between these subspaces is
illustrated in \fref{f:grids}.

It is also instructive to invert \eref{e:pqphys} and write the physical
{\coord}s $h_0$ and $\cosi$ in terms of $\AR$ and $\AL$:
\begin{equation}
  h_0 = \left(\sqrt{\AR} + \sqrt{\AL}\right)^2
  \quad\hbox{and}\quad
  \cosi = \frac{\sqrt{\AR}-\sqrt{\AL}}{\sqrt{\AR}+\sqrt{\AL}}
  \ ,
\end{equation}
which can be related to the CPF {\coord}s $\{\A^{\mudot}\}$ by
\begin{equation}
  \AR = \sqrt{(\Rx)^2+(\Ry)^2}
  \quad\hbox{and}\quad
  \AL = \sqrt{(\Lx)^2+(\Ly)^2}
  \ .
\end{equation}

\subsubsection{Root-radius {\Coord}s}

\label{s:coords-rootrad}

Finally, it is sometimes useful to define so-called root-radius
{\coord}s
\begin{subequations}
  \begin{align}
    \rR = \AR^{1/4} &= h_0^{1/4}\sqrt{\frac{1+\cosi}{2}}
    \\
    \rL = \AL^{1/4} &= h_0^{1/4}\sqrt{\frac{1-\cosi}{2}}
  \end{align}
\end{subequations}
with corresponding Cartesian {\coord}s
\begin{subequations}
  \begin{align}
    \xR = \rR\cos\tR
    \qquad&\hbox{and}\qquad
    \yR = \rR\sin\tR
    \\
    \xL = \rL\cos\tL
    \qquad&\hbox{and}\qquad
    \yL = \rL\sin\tL
  \end{align}
\end{subequations}
The relationship of these {\coord}s to the CPF {\coord} system is
\begin{subequations}
  \begin{align}
    \Rx = \rR^3\,\xR
    \qquad&\hbox{and}\qquad
    \Ry = \rR^3\,\yR
    \\
    \Lx = \rL^3\,\xL
    \qquad&\hbox{and}\qquad
    \Ly = \rL^3\,\yL
    \ ,
  \end{align}
\end{subequations}
and the physical parameters can be written
\begin{equation}
  h_0 = (\rR^2+\rL^2)^2
  \qquad\hbox{and}\qquad
  \cosi = \frac{\rR^2-\rL^2}{\rR^2+\rL^2}
  \ .
\end{equation}

\section[Applications]{Applications of the New {\Coord}s}
\label{s:apps}

\subsection[Jacobian]{Calculation of the Jacobian Between Physical and
  JKS {\Coord}s}

\label{s:apps-jacobian}

We can obtain the Jacobian determinant for the transformation between
the physical amplitude parameters
$\{h_0,\cosi\mbe\cos\iota,\psi,\phi_0\}$ and the JKS {\coord}s
$\{\A^1,\A^2,\A^3,\A^4\}$ by treating the transformation as a sequence
of transformations in which the {\coord}s are being transformed in
pairs.

First, we invert \eref{e:PQ} to obtain
\begin{subequations}
  \begin{align}
    \A^1 &= \Rx + \Lx
    \\
    \A^2 &= \Ry - \Ly
    \\
    \A^3 &= -\Ry - \Ly
    \\
    \A^4 &= \Rx - \Lx
    \ ,
  \end{align}
\end{subequations}
which produces the Jacobian determinants
\begin{equation}
  d\A^1\,d\A^4 = 2\,d\Rx\,d\Lx
  \quad\hbox{and}\quad
  d\A^2\,d\A^3 = 2\,d\Ry\,d\Ly
  \ .
\end{equation}

Next, since \eref{e:PQpolar} define $\{\AR,\tR\}$ as the polar
{\coord}s corresponding to the Cartesian {\coord}s
$\{\Rx,\Ry\}$, and likewise for $\{\AL,\tL\}$ and
$\{\Lx,\Ly\}$, the relevant Jacobian determinants are
\begin{equation}
  d\Rx\,d\Ry = \AR\,d\AR\,d\tR
  \quad\hbox{and}\quad
  d\Lx\,d\Ly = \AL\,d\AL\,d\tL
  \ .
\end{equation}

Finally, the identifications \eref{e:pqphys} lead to the Jacobian
matrices
\begin{equation}
  \begin{pmatrix}
    d\AR \\ d\AL
  \end{pmatrix}
  =
  \begin{pmatrix}
    \left(\frac{1+\cosi}{2}\right)^2 & h_0 \frac{1+\cosi}{2}
    \\
    \left(\frac{1-\cosi}{2}\right)^2 & - h_0 \frac{1-\cosi}{2}
  \end{pmatrix}
  \begin{pmatrix}
    dh_0 \\ d\cosi
  \end{pmatrix}
\end{equation}
and
\begin{equation}
  \begin{pmatrix}
    d\tR \\ d\tL
  \end{pmatrix}
  =
  \begin{pmatrix}
    2 & 1 \\ -2 & 1
  \end{pmatrix}
  \begin{pmatrix}
    d\psi \\ d\phi_0
  \end{pmatrix}
  \ ,
\end{equation}
whose determinants tell us
\begin{equation}
  d\AR\,d\AL = h_0\frac{1-\cosi^2}{4}\,dh_0\,d\cosi
  \quad\hbox{and}\quad
  d\tR\,d\tL = 4\,d\psi\,d\phi_0
  \ .
\end{equation}

To combine the effects of these three transformations, note from
\eref{e:pqphys} that
\begin{equation}
  \label{e:pqprod}
  \AR\AL = \left(h_0\frac{1-\cosi^2}{4}\right)^2
\end{equation}
and thus
\begin{multline}
  \label{e:Amu-phys}
  d\A^1\,d\A^2\,d\A^3\,d\A^4
  = 4\,d\Rx\,d\Ry\,d\Lx\,d\Ly
  \\
  = 4\AR\AL\,d\AR\,d\AL\,d\tR\,d\tL
  \\
  = 16\left(h_0\frac{1-\cosi^2}{4}\right)^3\,dh_0\,d\cosi\,d\psi\,d\phi_0
\end{multline}
which is the same as the form \eref{e:dAdphys} presented in
\cite{Prix09:_Bstat}.

Note that, using \eref{e:pqprod} we can rewrite \eref{e:Amu-phys} as
\begin{multline}
  dh_0\,d\cosi\,d\psi\,d\phi_0 = \frac{d\AR\,d\AL\,d\tR\,d\tL}{4\sqrt{\AR\AL}}
  \\
  = \frac{d\Rx\,d\Ry\,d\Lx\,d\Ly}
  {
    4
    \bigl[(\Rx)^2+(\Ry)^2\bigr]^{3/4}
    \bigr[(\Lx)^2+(\Ly)^2\bigr]^{3/4}
  }
  \ .
\end{multline}
If we recall the root-radius {\coord}s defined so that $\rR^2=\AR^{1/2}$ and
$\rL^2=\AL^{1/2}$, we have
\begin{equation}
  \frac{d\AR}{2\sqrt{\AR}} = 2\rR\,d\rR
  \qquad\hbox{and}\qquad
  \frac{d\AL}{2\sqrt{\AL}} = 2\rL\,d\rL
  \ ,
\end{equation}
so that
\begin{equation}
  \label{e:measure-rootrad}
  \begin{split}
    dh_0\,d\cosi\,d\psi\,d\phi_0 &= 4\,\rR\,d\rR\,d\tR\,\rL\,d\rL\,d\tL
    \\
    &= 4\,d\xR\,d\yR\,d\xL\,d\yL
    \ ,
  \end{split}
\end{equation}
i.e., the natural measure in physical {\coord}s is, up to a constant,
just the usual Lebesgue measure on a Cartesian space.

\subsection[Circular Polarization]{Nature of the {\Coord}
  Singularities for Circular Polarization}

\label{s:apps-circpol}

The volume element \eref{e:Amu-phys} has singularities in terms of the
physical {\coord}s for circular polarization, i.e., $\cosi=\pm 1$,
because the Jacobian
\begin{equation}
  \label{e:Jacobian}
  \begin{split}
    \mc{J}
    &= 2 \left(h_0\frac{1-\cosi^2}{2}\right)^3
    = 2\left(\Ap^2-\Ac^2\right)^{3/2} = 16\, (\AR\AL)^{3/2}
    \\
    &= 16\,
    \Bigl\{
    \Bigl[(\Rx)^2+(\Ry)^2\Bigr]
    \Bigr[(\Lx)^2+(\Ly)^2\Bigr]
    \Bigr\}^{3/4}
  \end{split}
\end{equation}
vanishes there.

\subsubsection{Right Circular Polarization ($\cosi=1$, i.e.,
  $\iota=0$)}

When $\cosi=1$, so that $\Ac=\Ap$, the polar amplitude {\coord}s
become $\AR=h_0$ and $\AL=0$, so the combination $\AR\AL$ vanishes, and the
amplitude parameters become
\begin{subequations}
  \label{e:rightcircpol}
  \begin{align}
    \A^1 = \A^4 = \Rx &= h_0\cosfpt \\
    \A^2 = -\A^3 = \Ry &= h_0\sinfpt \\
    \Lx = \Ly &= 0
  \end{align}
\end{subequations}
The waveform \eref{e:signal} is completely described by the amplitude
$h_0$ and the phase $\tR=\phi_0+2\psi$, exhibiting the well-known
degeneracy between $\psi$ and $\phi_0$ for circular polarization.

\subsubsection{Left Circular Polarization ($\cosi=-1$, i.e.,
  $\iota=\pi$)}

When $\cosi=-1$, so that $\Ac=-\Ap$, the polar amplitude {\coord}s
become $\AR=0$ and $\AL=h_0$, so the combination $\AR\AL$ vanishes, and the
amplitude parameters become
\begin{subequations}
  \label{e:leftcircpol}
  \begin{align}
    \Rx = \Ry &= 0 \\
    \A^1 = -\A^4 = \Lx &= h_0\cosfmt \\
    -\A^2 = -\A^3 = \Ly &= h_0\sinfmt
  \end{align}
\end{subequations}
The waveform \eref{e:signal} is completely described by the amplitude
$h_0$ and the phase $\tL=\phi_0-2\psi$, exhibiting the well-known
degeneracy between $\psi$ and $\phi_0$ for circular polarization.

\section{Integration techniques for the $\B$-statistic targeted
  search method}

\label{s:Bstat}

Prix and Krishnan \cite{Prix09:_Bstat} consider the case of a targeted
search, where the signal hypothesis $\mc{H}_s$ has known
phase-evolution parameters $\{\alpha,\delta,f_0,f_1,\ldots\}$ but
unknown amplitude parameters, obeying some prior probability
distribution $\pdf(\A|\mc{H}_s)$.\footnote{We will adopt the
  convention that $\A$ with no superscripts refers to an arbitrary set
  of {\coord}s on the four-dimensional amplitude parameter space,
  while $\{\A^\mu\}$, $\{\A^{\mudot}\}$, etc, refer to a specific set
  of {\coord}s.}  Given some observed data $\detV{x}$, they
calculate the Bayes factor
\begin{equation}
  \label{e:Bstatdef}
  \begin{split}
    \B(\detV{x})
    &= \frac{\pdf(\detV{x}|\mc{H}_s)}{\pdf(\detV{x}|\mc{H}_n)}
    = \frac{\int\pdf(\detV{x}|\mc{H}_s,\A)\,\pdf(\A|\mc{H}_s)\,d^4\!\A}
    {\pdf(\detV{x}|\mc{H}_n)}
    \\
    &= \int e^{\Lambda(\A;\detV{x})}\,\pdf(\A|\mc{H}_s)\,d^4\!\A
    \ ,
  \end{split}
\end{equation}
which they call the
$\B$-statistic, in contrast with the $\F$-statistic, which is
the maximum log-likelihood ratio
\begin{equation}
  \F(\detV{x})
  = \max_{\A} \ln\frac{\pdf(\detV{x}|\mc{H}_s,\A)}{\pdf(\detV{x}|\mc{H}_n)}
  = \max_{\A} \Lambda(\A;\detV{x})
  \ .
\end{equation}

\subsection{Form of the Log-Likelihood Ratio}
\label{s:Bstat-loglike}

The log-likelihood ratio can be written in the form\footnote{As
  before, we use the Einstein summation convention to imply sums over
  $\mu$ and $\nu$ as appropriate.}
\cite{krolak04:_optim_lisa,WhelanPrix08:_MLDC1B}
\begin{equation}
  \Lambda(\{\A^\mu\};\detV{x})
  = \A^\mu x_\mu - \frac{1}{2}\A^\mu \M_{\mu\nu} \A^\nu
\end{equation}
where we defined
\begin{equation}
  x_\mu \equiv \scalar{\detV{x}}{\detV{h}_\mu}
\ ,\quad\text{and}\quad
  \M_{\mu\nu} \equiv \scalar{\detV{h}_\mu}{\detV{h}_\nu} 
\ , \label{eq:Mmunu}
\end{equation}
in terms of the standard scalar product $\scalar{\cdot}{\cdot}$ defined in
\eref{eq:def-scalar}, the strain data $x$ and
the four scalar basis waveforms $h_\mu$, which are the detector's
response to the four GW tensor functions $\tens{h}_\mu$ according to
\eref{e:response}.
As shown in \aref{app:metric}, the matrix
$\{\M_{\mu\nu}\}$ is explicitly found to have the form
\begin{equation}
  \label{e:Mmunu}
  \{\M_{\mu\nu}\}
  =
  \begin{pmatrix}
    A &  C &  0 & E \\
    C &  B & -E & 0 \\
    0 & -E &  A & C \\
    E &  0 &  C & B
  \end{pmatrix}
  \ .
\end{equation}
In the long-wavelength limit, $E=0$; it is non-zero only in the regime
where the finite size of the detector is important, and the simple
response tensor \eref{e:dLWL} is replaced by a complex
frequency-dependent expression.

Since the new amplitude {\coord}s
$\{\A^{\mudot}\}
$ are linear combinations of the $\{\A^\mu\}$, we can also write the
log-likelihood ratio as a quadratic in those {\coord}s:
\begin{equation}
  \label{e:loglike}
  \Lambda(\{\A^{\mudot}\};\detV{x})
  =  \A^{\mudot} x_{\mudot}
  - \frac{1}{2}
  \A^{\mudot} \M_{\mudot\nudot} \A^{\nudot}
  \ ,
\end{equation}
with
\begin{equation}
  x_{\mudot} = \scalar{x}{h_{\mudot}}
\ ,\quad\text{and}\quad
  \M_{\mudot\nudot} = \scalar{h_{\mudot}}{h_{\nudot}} 
\ ,
\end{equation}
in analogy to \eref{eq:Mmunu}, and the transformed matrix is
found to have the form
\begin{equation}
  \label{e:Mdot}
  \{\M_{\mudot\nudot}\}
  =
  \begin{pmatrix}
    I &  0 &  L & -K \\
    0 &  I &  K &  L \\
    L &  K &  J &  0 \\
   -K &  L &  0 &  J
  \end{pmatrix}
  \ ,
\end{equation}
with the explicit matrix elements given in \aref{app:metric}, and in
the long-wavelength limit we have $I=J$.

In terms of the data vector $\{x_{\mudot}\}$ (whose explicit form
is given in \aref{app:metric}), the linear part of the log-likelihood
ratio is
\begin{equation}
  \label{e:loglikelin}
  \A^{\mudot} x_{\mudot}
  = \AR (x_{\onedot}\cos\tR+x_{\twodot}\sin\tR)
  + \AL (x_{\tredot}\cos\tL+x_{\fordot}\sin\tL)
  \ .
\end{equation}

The quadratic part of the log-likelihood can be written in the
$\{\A^{\mudot}\}$ {\coord}s as
\begin{equation}
  \label{e:loglikequad}
  \begin{split}
  \A^{\mudot} \M_{\mudot\nudot}& \A^{\nudot}
  = I [(\Rx)^2+(\Ry)^2] + J [(\Lx)^2+(\Ly)^2]
  \\
  &+ 2K [-\Rx\Ly + \Ry\Lx]
  + 2L [\Rx\Lx + \Ry\Ly]
  \\
  =\ & I \AR^2 + J \AL^2
  \\
  &+ 2\AR\AL
  \left[
    K\sin(\tR-\tL) + L\cos(\tR-\tL)
  \right]
  \ .
  \end{split}
\end{equation}
Note that this depends upon the angular {\coord}s only in the
combination $\tR-\tL=4\psi$, and is independent of
$\tR+\tL=2\phi_0$.

Because the amplitude parameters $\{\A^{\mudot}\}$ which maximize the
log-likelihood ratio $\Lambda(\{\A^{\mudot}\};\detV{x})$ are given
by
\begin{equation}
  \MLE{\A}^{\mudot}(\detV{x}) = \M^{\mudot\nudot} x_{\nudot}
  \ ,
\end{equation}
where $\{\M^{\mudot\nudot}\}$ is the matrix inverse of
$\{\M_{\mudot\nudot}\}$, and the maximum of the log-likelihood ratio
is the $\F$-statistic
\begin{equation}
  \F(\detV{x}) =  \frac{1}{2}x_{\mudot} \M^{\mudot\nudot} x_{\nudot}
  = \frac{1}{2}\MLE{\A}^{\mudot}(\detV{x})
  \M_{\mudot\nudot} \MLE{\A}^{\nudot}(\detV{x})
  \ ,
\end{equation}
it is convenient to write the log-likelihood ratio as
\begin{equation}
  \label{e:loglikeF}
  \Lambda(\{\A^{\mudot}\};\detV{x})
  = \F(\detV{x}) - \frac{1}{2}
  [\A^{\mudot}-\MLE{\A}^{\mudot}(\detV{x})]
  \M_{\mudot\nudot}
  [\A^{\nudot}-\MLE{\A}^{\nudot}(\detV{x})]
  \ .
\end{equation}

\subsection{Integration in CPF {\coord}s}
\label{s:Bstat-sigcoord}

Since the log-likelihood ratio $\Lambda(\A;\detV{x})$
\eref{e:loglikeF} is quadratic in
$\{\A^{\mudot}\}
$ (just as it
is in $\{\A^\mu\}$) we can do the Gaussian integral, for the case of
the unphysical $\F$-statistic prior \eref{e:Fstatprior}, as shown
in \cite{Prix09:_Bstat}:
\begin{equation}
  \begin{split}
    \B(\detV{x})
    &= \int e^{\Lambda(\A;\detV{x})}\,\pdf(\A|\Hf)\,d^4\!\A
    \\
    &\propto \int
    e^{
      \A^{\mudot} x_{\mudot}
      - \frac{1}{2}
      \A^{\mudot} \M_{\mudot\nudot} \A^{\nudot}
    }
    \,d\A^{\onedot}\,d\A^{\twodot}\,d\A^{\tredot}\,d\A^{\fordot}
    \\
    &= e^{\F(\detV{x})}
    \int
    e^{
      - \frac{1}{2}\Delta\A^{\mudot}
      \M_{\mudot\nudot}
      \Delta\A^{\nudot}
    }
    \,d\A^{\onedot}\,d\A^{\twodot}
    \,d\A^{\tredot}\,d\A^{\fordot}
    \\
    &\propto e^{\F(\detV{x})}
    \ .
  \end{split}
\end{equation}
If, however, an isotropic prior\eref{e:physpdf} is used, so that
\begin{multline}
  \label{e:pdfsignal}
  \pdf(\{\A^{\mudot}\}|\Hi)
  = \pdf(\Rx,\Ry,\Lx,\Ly|\Hi)
  \\
  = \frac{4}{\mc{J}}\pdf(h_0,\cosi,\psi,\phi_0|\Hi)
  = \frac{2}{\pi^2}\frac{\pdf(h_0|\Hi)}{\mc{J}}
  \ ,
\end{multline}
where $\mc{J}$ is the Jacobian determinant specified
in \eref{e:Jacobian}.  Then if we define
\begin{equation}
  \alpha(\A) = \ln \frac{\pdf(h_0|\Hi)}{\mc{J}}
  \ ,
\end{equation}
the $\B$-statistic integral can be written as
\begin{equation}
  \label{e:Bstat-with-measure}
  \B(\detV{x})
  \propto
  e^{\F(\detV{x})}
  \int
  e^{
    -\frac{1}{2} \Delta\A^{\mudot}\M_{\mudot\nudot}\Delta\A^{\nudot}
    + \alpha(\A)
  }
  \,d\A^{\onedot}\,d\A^{\twodot}
  \,d\A^{\tredot}\,d\A^{\fordot}
  \ .
\end{equation}
One possible approach would be to Taylor expand $\alpha(\A)$ about the
maximum-likelihood point $\MLE{\A}$,
\begin{equation}
  \alpha(\A) = \MLE{\alpha} + \MLE{\alpha}_{\mudot}\Delta\A^{\mudot}
  + \frac{1}{2}
  \MLE{\alpha}_{\mudot\nudot}\Delta\A^{\mudot}\Delta\A^{\nudot}
  + \mc{O}([\Delta\A]^3)
  \ ,
\end{equation}
where we have defined the expansion coefficients
\begin{subequations}
  \begin{align}
    \MLE{\alpha} &= \alpha(\MLE{\A})
    \\
    \MLE{\alpha}_{\mudot}
    &=
    \left.
      \frac{\partial\alpha}{\partial\A^{\mudot}}
    \right\rvert_{\A=\MLE{\A}}
    \\
    \MLE{\alpha}_{\mudot\nudot}
    &=
    \left.
      \frac{\partial^2\alpha}{\partial\A^{\mudot}\partial\A^{\nudot}}
    \right\rvert_{\A=\MLE{\A}} \, .
  \end{align}
\end{subequations}
(This was the method used in Cohen et al.~\cite{2010CQGra..27r5012C}
for approximating the analog of the $\B$-statistic for the case of
GW bursts from cosmic strings.)  We could then approximate the
integral as Gaussian, obtaining the result
\begin{equation}
  \begin{split}
    \B(\detV{x})
    \propto&
    \left(\det\{\mc{N}_{\mudot\nudot}(\detV{x})\}\right)^{-1/2}
    \\
    &\times
    \exp
    \left(
      \F(\detV{x})+\MLE{\alpha}(\detV{x})\
      + \frac{1}{2}
      \MLE{\alpha}_{\mudot}(\detV{x})
      \mc{N}^{\mudot\nudot}(\detV{x})
      \MLE{\alpha}_{\nudot}(\detV{x})
    \right)
    \ ,
  \end{split}
\end{equation}
where we have defined the matrix
\begin{equation}
  \mc{N}_{\mudot\nudot}(\detV{x}) = \M_{\mudot\nudot}
  - \MLE{\alpha}_{\mudot\nudot}(\detV{x})
\end{equation}
and its inverse $\mc{N}^{\mudot\nudot}$ (so that
$\mc{N}^{\mudot\nudot}\mc{N}_{\nudot\lamdot}=\delta_{\lam}^{\mu}$).
However, this approximation can only be valid if the matrix
$\mc{N}^{\mudot\nudot}$ is positive definite, so that the point
$\A^{\mudot}=\MLE{\A}^{\mudot}+\mc{N}^{\mudot\nudot}\MLE{\alpha}_{\nudot}$
is a maximum of the integrand in \eref{e:Bstat-with-measure}.  We
will show that that is not in general true by calculating the explicit
form of $\MLE{\alpha}_{\mudot\nudot}$.

We limit attention to the simple case of a uniform prior on $h_0$,
$\pdf(h_0|\Hi)=\text{const}$.  In that case,
\begin{multline}
  \label{e:alphadef}
    \alpha(\A) = -\ln\mc{J} + \text{const}
    = -\frac{3}{2}\ln(\AR\AL) + \text{const}
    \\
    = -\frac{3}{4}\ln\left([\Rx]^2+[\Ry]^2\right)
    -\frac{3}{4}\ln\left([\Lx]^2+[\Ly]^2\right)
    + \text{const}
    \ ,
\end{multline}
and we can calculate the unique non-vanishing derivatives as
\begin{equation}
  \frac{\partial\alpha}{\partial\Rx}
  = -\frac{3}{2}\frac{\Rx}{[\Rx]^2+[\Ry]^2}
\end{equation}
and
\begin{subequations}
  \begin{align}
    \frac{\partial^2\alpha}{(\partial\Rx)^2}
    &= -\frac{3}{2}
    \frac{[\Ry]^2-[\Rx]^2}{([\Rx]^2+[\Ry]^2)^2}
    = \frac{3}{2}\frac{\cos 2\tR}{\AR^2}
    \\
    \frac{\partial^2\alpha}{\partial\Rx\partial\Ry}
    &= \frac{3}{2}
    \frac{2\Rx\Ry}{([\Rx]^2+[\Ry]^2)^2}
    = \frac{3}{2}\frac{\sin 2\tR}{\AR^2}
    \ ,
  \end{align}
\end{subequations}
with the derivatives with respect to the other $\{\A^{\mudot}\}$
following by inspection.  The resulting matrix is
\begin{equation}
  \MLE{\alpha}_{\mudot\nudot}
  = \frac{3}{2}
  \begin{pmatrix}
    \frac{\cos 2\MLEtR}{\MLEAR^2}
    & \frac{\sin 2\MLEtR}{\MLEAR^2}
    & 0 & 0
    \\
    \frac{\sin 2\MLEtR}{\MLEAR^2}
    & -\frac{\cos 2\MLEtR}{\MLEAR^2}
    & 0 & 0
    \\
    0 & 0
    & \frac{\cos 2\MLEtL}{\MLEAL^2}
    & \frac{\sin 2\MLEtL}{\MLEAL^2}
    \\
    0 & 0
    & \frac{\sin 2\MLEtL}{\MLEAL^2}
    & -\frac{\cos 2\MLEtL}{\MLEAL^2}
  \end{pmatrix}
  \ .
\end{equation}
Now, if the data happen to be such that the maximum likelihood
estimates of the amplitude parameters $\MLE{\A}$ correspond to right-
or left-circular polarization, then the parameter $\MLEAR$ or
$\MLEAL$, respectively, will be small.  Since the metric
$\{\M_{\mudot\nudot}\}$ is determined by the observing geometry
and the noise level, and not the realization of the data, it can
always happen that $\MLEAR$ or $\MLEAL$ is small enough that two of
the eigenvalues of $\mc{N}_{\mudot\nudot}(\detV{x}) =
\M_{\mudot\nudot}
-\MLE{\alpha}_{\mudot\nudot}(\detV{x})$ will be approximately
equal to the corresponding eigenvalues of
$-\MLE{\alpha}_{\mudot\nudot}(\detV{x})$, which will be the
eigenvalues of the matrix
\begin{equation}
  \frac{3}{2\MLEAR^2}
  \begin{pmatrix}
    -\cos 2\MLEtR & -\sin 2\MLEtR
    \\
    -\sin 2\MLEtR & \cos 2\MLEtR
  \end{pmatrix}
  \ ,
\end{equation}
which are $\pm\frac{3}{2\AR^2}$ (or the corresponding expression
involving $\AL$, in the case of left circular polarization).  Since
these two eigenvalues have opposite signs,
$\{\mc{N}_{\mudot\nudot}\}$ is not a positive definite matrix,
the point $\A^{\mudot}
=\MLE{\A}^{\mudot}+\mc{N}^{\mudot\nudot}\MLE{\alpha}_{\nudot}$
is a saddle point rather than a maximum, and the Gaussian
approximation for the integral \eref{e:Bstat-with-measure} fails.

One issue with this approach is that the maximum likelihood point is a
stationary point of $\Lambda(\A;\detV{x})$ rather than
$\Lambda(\A;\detV{x})+\alpha(\A)$, and we should consider expanding
about the maximum of $\Lambda(\A;\detV{x})+\alpha(\A)$.  In fact,
$\Lambda(\A;\detV{x})+\alpha(\A)$ has no global maximum, as
examination of \eref{e:alphadef} shows that
$\alpha(\A)\rightarrow+\infty$ as $\AR$ or $\AL$ goes to zero.  The best
we can hope for is a local maximum when
\begin{equation}
  \frac{\partial}{\partial\A^{\mudot}}
  \left[\Lambda(\A;\detV{x})+\alpha(\A)\right]
  = x_{\mudot}
  - \M_{\mudot\nudot} \A^{\nudot}
  + \alpha_{\mudot}(\A) = 0
  \ .
\end{equation}
This local maximum can fail to exist even when the matrix
$\{\mc{N}_{\mudot\nudot}\}$ is positive definite, and in any event,
the Gaussian integral would only approximate the area under the local
maximum, not the contribution from the integrable singularity at $\AR=0$
and $\AL=0$.  This is examined in further detail in
\sref{s:explicit-sigcoord}.

\subsection{Integration in root-radius {\coord}s}
\label{s:Bstat-rootrad}

As we've seen in \sref{s:Bstat-sigcoord}, while the log-likelihood
ratio is quadratic in CPF (or JKS) {\coord}s, the integrand of the
$\B$-statistic integral arising from an isotropic prior
\eref{e:physpdf} contains a {\coord} singularity which prevents the
integrand from being approximated by a Gaussian.  If we focus
attention on the constant-$h_0$ prior
\begin{equation}
  \pdf(h_0,\cosi,\psi,\phi_0|\Hi) = \text{const}
  \ ,
\end{equation}
the measure of the integral will be constant not only in physical
{\coord}s $\{h_0,\cosi,\psi,\phi_0\}$ but also in the root-radius
Cartesian {\coord}s $\{\xR,\yR,\xL,\yL\}$ defined in
\sref{s:coords-rootrad} [see \eref{e:measure-rootrad}].  The integral
will not be a Gaussian, since the log-likelihood ratio will no longer
be quadratic in these {\coord}s, but it will remain non-singular and
have a single maximum at the point $\A=\MLE{\A}$.  Thus we can write
the integral as
\begin{equation}
  \B(\detV{x})
  \propto
  \int_{-\infty}^{\infty}
  \int_{-\infty}^{\infty}
  \int_{-\infty}^{\infty}
  \int_{-\infty}^{\infty}
  e^{\Lambda(\A;\detV{x})}
  \,d\xR\,d\yR\,d\xL\,d\yL
\end{equation}
and attempt to Taylor expand the log of the integrand,
$\Lambda(\A;\detV{x})$, about its maximum.  Writing
\begin{equation}
  \{\A^\alpha\} = \{\xR, \yR, \xL, \yL\}
  \ ,
\end{equation}
the expansion is
\begin{equation}
  \begin{split}
    \Lambda(\A;\detV{x})
    = & \Lambda(\MLE{\A};\detV{x})
    + \left.\frac{\partial\Lambda}{\partial\A^\alpha}\right\rvert_{\A=\MLE{A}}
    (\A^\alpha-\MLE{\A}^\alpha)
    \\
    & + \frac{1}{2}
    \left.
      \frac{\partial^2\Lambda}{\partial\A^\alpha\partial\A^\beta}\right\rvert_{\A=\MLE{A}}
    (\A^\alpha-\MLE{\A}^\alpha)(\A^\beta-\MLE{\A}^\beta)
    \\
    &+ \mc{O}([\A-\MLE{\A}]^3)
    \ .
  \end{split}
\end{equation}
Since $\MLE{\A}$ is the maximum-likelihood point,
$\Lambda(\MLE{\A};\detV{x})=\F(\detV{x})$ and
$\left.\frac{\partial\Lambda}{\partial\A^\alpha}\right\rvert_{\A=\MLE{A}}=0$.
If we use  \eref{e:loglikeF} to calculate
\begin{equation}
  \frac{\partial^2\Lambda}{\partial\A^\alpha\partial\A^\beta}
  =
  \frac{\partial^2\A^{\mudot}}{\partial\A^\alpha\partial\A^\beta}
  \M_{\mudot\nudot}
  [\A^{\nudot}-\MLE{\A}^{\nudot}(\detV{x})]
  + \frac{\partial\A^{\mudot}}{\partial\A^\alpha}\M_{\mudot\nudot}
  \frac{\partial\A^{\nudot}}{\partial\A^\beta}
  \ ,
\end{equation}
we see that
\begin{equation}
  \left.
    \frac{\partial^2\Lambda}
    {\partial\A^\alpha\partial\A^\beta}
  \right\rvert_{\A=\MLE{A}}
  =
  \left.
    \frac{\partial\A^{\mudot}}{\partial\A^\alpha}
  \right\rvert_{\A=\MLE{A}}
  \M_{\mudot\nudot}
  \left.
    \frac{\partial\A^{\nudot}}{\partial\A^\beta}
  \right\rvert_{\A=\MLE{A}}
  \ .
\end{equation}
Thus
\begin{multline}
  \Lambda(\A;\detV{x})
  \\
  = \F(\detV{x})
  + \frac{1}{2}
  \left.
    \frac{\partial\A^{\mudot}}{\partial\A^\alpha}
  \right\rvert_{\A=\MLE{A}}
  \M_{\mudot\nudot}
  \left.
    \frac{\partial\A^{\nudot}}{\partial\A^\beta}
  \right\rvert_{\A=\MLE{A}}
  (\A^\alpha-\MLE{\A}^\alpha)(\A^\beta-\MLE{\A}^\beta)
  \\
  + \mc{O}([\A-\MLE{\A}]^3)
  \ ;
\end{multline}
if we keep only the quadratic piece, we get the approximate Gaussian
integral
\begin{equation}
  \begin{split}
    \ln\B(\detV{x})
    &\approx
    \F(\detV{x})
    - \frac{1}{2}
    \ln\det
    \left\{
      \left.
        \frac{\partial^2\Lambda}
        {\partial\A^\alpha\partial\A^\beta}
      \right\rvert_{\A=\MLE{A}(\detV{x})}
    \right\}
    + \text{const}
    \\
    &\approx
    \F(\detV{x})
    -
    \ln\det
    \left\{
      \left.
        \frac{\partial\A^{\mudot}}{\partial\A^\alpha}
      \right\rvert_{\A=\MLE{A}(\detV{x})}
    \right\}
    + \text{const}
    \ ,
  \end{split}
\end{equation}
where we have absorbed the term
$-\frac{1}{2}\det\left\{\M_{\mudot\nudot}\right\}$ into the
constant since it does not depend on the data.

The determinant
\begin{equation}
  \det
  \left\{
    \frac{\partial\A^{\mudot}}{\partial\A^\alpha}
  \right\}
  = 16 (\AR\AL)^{3/2}
\end{equation}
has already been calculated, since it's the Jacobian for the
transformation between the {\coord}s $\{\A^\alpha\} = \{\xR, \yR, \xL,
\yL\}$ and $\{\A^{\mudot}\}=\{\Rx,\Ry,\Lx,\Ly\}$.
This means that
\begin{equation}
  \label{e:Bgauss}
  \ln\B(\detV{x})
  \approx
  \F(\detV{x})
  - \frac{3}{2}
  \ln
  \left(
    \MLEAR(\detV{x})\,\MLEAL(\detV{x})
  \right)
  + \text{const}
  \ .
\end{equation}
This approximate correction factor in $\B(\detV{x})$ has a familiar
form: it's the Jacobian appearing in the Gaussian integral in CPF
{\coord}s, evaluated at the maximum likelihood point.  The
approximation again breaks down if the maximum likelihood point is too
close to circular polarization, i.e., if $\MLEAR(\detV{x})$ or
$\MLEAL(\detV{x})$ is close to zero.  It's easy to see why this is
the case: the log-likelihood-ratio $\Lambda(\A;\detV{x})$ has terms
proportional to $(\AR-\MLEAR(\detV{x}))^2$ and
$(\AL-\MLEAL(\detV{x}))^2$; if e.g., $\MLEAR(\detV{x})=0$, the
first term becomes $\AR^2=\rR^8=(\xR^2+\yR^2)^4$, and
$\Lambda(\A;\detV{x})$ cannot be approximated as quadratic in $\xR$
and $\yR$, since the second derivatives at the maximum likelihood
point vanish.  The resulting Gaussian is infinitely wide, leading to
the divergence of the approximated integral.  We examine where the
Gaussian approximation breaks down as a function of
$\MLEAR(\detV{x})$ and $\MLEAL(\detV{x})$ in
\sref{s:explicit-rootrad}.

\subsection{Integration in physical {\coord}s}

\label{s:Bstat-phys}

Continuing our consideration of the $\B$-statistic integral in the
case of a prior distribution uniform in the physical {\coord}s
$\{h_0,\cosi,\psi,\phi_0\}$, we turn to integration in the physical
{\coord}s themselves.  The measure of the integral is again constant,
while the log-likelihood ratio is more complicated.  By examining the
functional form of the integrand, we can see which integrals can be
performed exactly and which must be approximated or evaluated
numerically.  Using the explicit forms in \sref{s:Bstat-loglike}, and
keeping in mind the forms \eref{e:pqphys} of $\AR$ and $\AL$, we see that
\eref{e:loglikequad} is independent of $\phi_0$ and proportional to
$h_0^2$, so has the form
\begin{subequations}
  \label{e:loglikeh0phi0}
  \begin{equation}
    \A^{\mudot} \M_{\mudot\nudot} \A^{\nudot}
    = h_0^2 [\gamma(\cosi,\psi)]^2
  \end{equation}
  while \eref{e:loglikelin} is proportional to $h_0$ and depends on
  trigonometric functions of $\tR=\phi_0+2\psi$ and $\tL=\phi_0-2\psi$;
  it can thus be written
  \begin{equation}
    \A^{\mudot} x_{\mudot}
    = h_0\,\omega(\detV{x};\cosi,\psi)
    \cos(\phi_0-\varphi_0(\detV{x};\cosi,\psi))
    \ .
  \end{equation}
\end{subequations}
Inserting this form of the log likelihood into \eref{e:Bstatdef} and
assuming the isotropic prior \eref{e:physpdf} gives us
\begin{multline}
  \B \propto
  \int_{-1}^{1}
  \int_{-\pi/4}^{\pi/4}
  \int_{0}^{\infty}
  \pdf(h_0|\Hi)
  \, e^{-\frac{1}{2}h_0^2[\gamma(\cosi,\psi)]^2}
  \\
  \int_{0}^{2\pi}
  \exp
  \left\{
    h_0\,\omega(\detV{x};\cosi,\psi)
    \cos[\phi_0-\varphi_0(\detV{x};\cosi,\psi)]
  \right\}
  \,d\phi_0
  \\
  dh_0
  \,d\psi
  \,d\cosi
  \ .
\end{multline}
The integration over $\phi_0$ can be performed by using the
Jacobi-Anger expansion to show that
$\int_0^{2\pi}e^{x\,\cos\phi}\,d\phi = 2\pi \, I_0(x)$, where $I_0(x)
= J_0(ix)$ is the modified Bessel function of the first kind (cf
\cite{abramowitz64:_handb_mathem_funct}). This results in
\begin{multline}
  \B \propto
  \int_{-1}^{1}
  \int_{-\pi/4}^{\pi/4}
  \int_{0}^{\infty}
  \pdf(h_0|\Hi)
  \, e^{-\frac{1}{2}h_0^2[\gamma(\cosi,\psi)]^2}
  \\
  I_0(h_0\,\omega(\detV{x};\cosi,\psi))
  \,dh_0
  \,d\psi
  \,d\cosi
  \ .
\end{multline}
If we once again consider the simple case of a prior which is uniform
in $h_0$ over the range of interest, we can use the identity
\begin{multline}
  \label{eq:74}
  \int_0^\infty e^{-a^2 t^2}\,I_\nu(b t)\, dt
  = \frac{\pi^{1/2}}{2 a}\, e^{b^2/8a^2} \,
  I_{\frac{\nu}{2}}\left(\frac{b^2}{8 a^2} \right)
  ,
  \\
  \Real(\nu)>-1, \Real(a^2)>0
\end{multline}
(see Eq.~11.4.31 in \cite{abramowitz64:_handb_mathem_funct}) to
perform the $h_0$ integral analytically as well, leaving a
two-dimensional integral for the $\B$-statistic:
\begin{equation}
  \label{e:approxBstat}
  \B \propto
  \int_{-1}^{1}
  \int_{-\pi/4}^{\pi/4}
  \frac{I_0(\xi(\detV{x};\cosi,\psi))\,e^{\xi(\detV{x};\cosi,\psi)}}
  {\gamma(\cosi,\psi)}
  \,d\psi
  \,d\cosi
  \ ,
\end{equation}
where
\begin{equation}
  \xi(\detV{x};\cosi,\psi)
  = \frac{[\omega(\detV{x};\cosi,\psi)]^2}{4[\gamma(\cosi,\psi)]^2}
  \ .
\end{equation}
Further approximation and/or numerical evaluation techniques, which
are beyond the scope of this paper, can be applied to the expression
\eref{e:approxBstat}.

\section{Explicit evaluation of $\B$-statistic integral}
\label{s:explicit}

\subsection{Exact Solution in CPF-polar {\Coord}s}

To get a more concrete sense of when the various approximations
described in the previous sections break down, we consider a special
case in which the integral defining the $\B$-statistic with the
prior pdf \eref{e:physpdf} can be explicitly evaluated.  This occurs
when we assume
\begin{subequations}
  \begin{gather}
    A = B = \frac{1}{2\hdet^2}
    \\
    C = E = 0
\ ,
  \end{gather}
\end{subequations}
so that \eref{e:Mdot} becomes
\begin{equation}
  \label{e:Mdiag}
  \M_{\mudot\nudot} = \hdet^{-2} \, \delta_{\mudot\nudot}
\end{equation}
and the log-likelihood ratio is
\begin{equation}
  \Lambda(\A;\detV{x})
  = \LambdaR(\AR,\tR;\MLEAR,\MLEtR) + \LambdaL(\AL,\tL;\MLEAL,\MLEtL)
\end{equation}
where
\begin{subequations}
  \begin{align}
    \LambdaR(\AR,\tR;\MLEAR,\MLEtR)
    &=
    \frac{1}{2}\frac{\AR^2}{\hdet^2}
    - \frac{\AR\MLEAR}{\hdet^2}\cos(\tR-\MLEtR)
    \\
    \LambdaL(\AL,\tL;\MLEAL,\MLEtL)
    &=
    \frac{1}{2}\frac{\AL^2}{\hdet^2}
    - \frac{\AL\MLEAL}{\hdet^2}\cos(\tL-\MLEtL)
  \end{align}
\end{subequations}
and likewise
\begin{subequations}
  \label{e:Fstatexplicit}
\begin{equation}
  \F(\detV{x}) = \FsR(\MLEAR) + \FsL(\MLEAL)
\end{equation}
where
\begin{equation}
  \FsR(\MLEAR) = \frac{1}{2}\frac{\MLEAR^2}{\hdet^2}
  \qquad\hbox{and}\qquad
  \FsL(\MLEAL) = \frac{1}{2}\frac{\MLEAL^2}{\hdet^2}
  \ .
\end{equation}
\end{subequations}
In these expressions the observed data $\detV{x}$ manifest
themselves in the maximum-likelihood values
$\{\MLEAR,\MLEAL,\MLEtR,\MLEtL\}$.  We have suppressed the
$\detV{x}$ dependence in the interest of simplifying the notation.

\begin{figure}
  \begin{center}
    \includegraphics[width=\columnwidth]{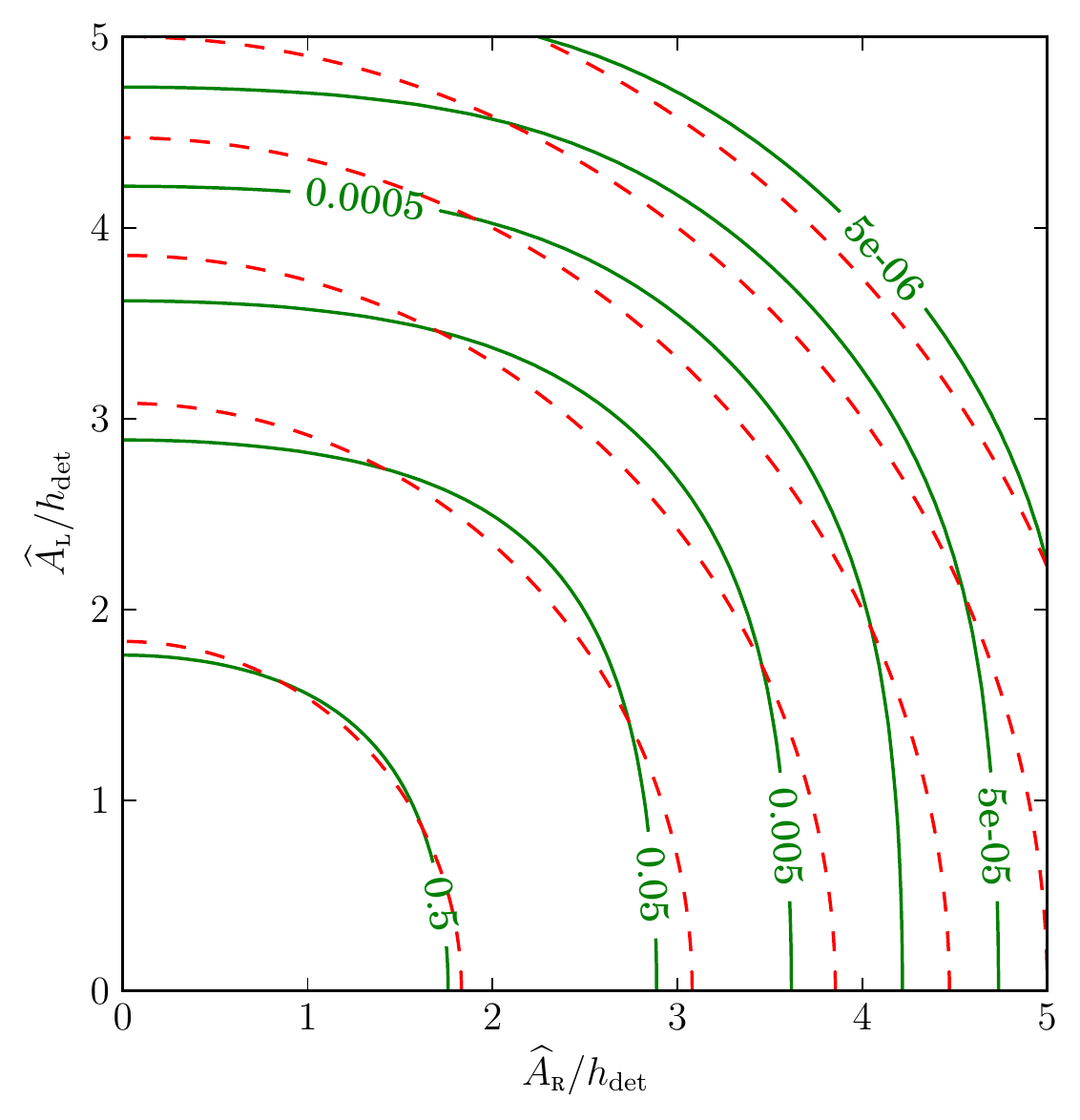}
  \end{center}
  \caption{Comparison of the $\B$ (solid) and $\F$ (dashed)
    statistics at equal false-alarm probabilities.  In the simple case
    of a diagonal amplitude parameter metric \eref{e:Mdiag}, we can
    explicitly evaluate the $\B$-statistic via
    \eref{e:Bstatexplicit} and the $\F$-statistic via
    \eref{e:Fstatexplicit}.  Because a prior distribution constant in
    the physical {\coord}s $\{h_0,\cosi,\psi,\phi_0\}$ weights
    circular polarization ($\AR$ or $\AL$ small) more heavily does than a
    prior uniform in the signal amplitudes $\{\A^{\mudot}\}$, we find
    that nearly circularly-polarized signals produce a $\B$
    statistic value more significant than their $\F$-statistic
    value, compared to nearly linearly-polarized signals (for which
    $\AR$ and $\AL$ are comparable).}
  \label{f:detectioncontours}
\end{figure}

The $\B$-statistic is then
\begin{equation}
  \B(\detV{x}) = \BR(\MLEAR)\BL(\MLEAL)
  \ ,
\end{equation}
where
\begin{equation}
  \label{e:explicitintegral}
  \frac{\BR(\MLEAR)}{\BR(0)}
  = C
  \int_{0}^{2\pi} \int_{0}^{\infty}
  e^{\LambdaR(\AR,\tR;\MLEAR,\MLEtR)}
  \AR^{-1/2}\,d\AR\,d\tR
  \ ,
\end{equation}
with a similar expression for $\BL(\MLEAL)/\BL(0)$.
We have put aside the question of normalization by writing an
expression for $\BR(\MLEAR)/\BR(0)$ and defining
\begin{equation}
  C = \frac{1}{2^{1/4}\pi\hdet^{1/2}\Gamma(1/4)}
  \ .
\end{equation}
We now demonstrate the explicit evaluation of the integral.  We
evaluate the $\tR$ integral as follows:
\begin{multline}
    \frac{\BR(\MLEAR)}{\BR(0)}
    \\
    = C
    \int_{0}^{2\pi} \int_{0}^{\infty}
    e^{\AR\MLEAR\cos(\tR-\MLEtR)/\hdet^2}
    \frac{e^{-\AR^2/2\hdet^2}}{\AR^{1/2}}
    \,d\AR\,d\tR
    \\
    = 2\pi A
    e^{-\MLEAR^2/2\hdet^2}
    \int_{0}^{\infty}
    \frac{e^{-\AR^2/2\hdet^2}}{\AR^{1/2}}
    \, I_0\left(\frac{\AR\MLEAR}{2\hdet^2}\right)
    \,d\AR
    \ ,
\end{multline}
where we have used the Jacobi-Anger expansion
\cite{abramowitz64:_handb_mathem_funct} and $I_0(x)=J_0(ix)$ is the
modified Bessel function of the first kind.  The $\AR$ integral can
also be done analytically, using identity (11.4.28) of
\cite{abramowitz64:_handb_mathem_funct}, with $a=2^{-1/2}\hdet^{-1}$,
$b=i\,\MLEAR\,\hdet^{-2}$, $\mu=1/2$, and $\nu=0$ to give
\begin{equation}
  \label{e:hypersoln}
  \frac{\BR(\MLEAR)}{\BR(0)}
  = {}_1F_1\left(\frac{1}{4},1,\frac{\MLEAR^2}{2\hdet^2}\right)
  \ ,
\end{equation}
where ${}_1F_1(a,b,z)=M(a,b,z)$ is the confluent hypergeometric
function.  Note that ${}_1F_1(a,b,0)=1$ by identity (13.5.5) of
\cite{abramowitz64:_handb_mathem_funct}.  The overall detection
statistic is thus
\begin{equation}
  \label{e:Bstatexplicit}
  \frac{\B(\detV{x})}{\B(\detVm{0})}
  = {}_1F_1\left(\frac{1}{4},1,\frac{\MLEAR^2}{2\hdet^2}\right)
  \ {}_1F_1\left(\frac{1}{4},1,\frac{\MLEAL^2}{2\hdet^2}\right)
  \ .
\end{equation}
In \fref{f:detectioncontours} we illustrate the difference between
$\B$ and $\F$ as detection statistics by plotting, versus
$\MLEAR(\detV{x})$ and $\MLEAL(\detV{x})$, surfaces of constant
$\B$ and $\F$, at the same set of false-alarm probabilities.

\subsection{Comparison to Root-Radius Gaussian Approximation}

\label{s:explicit-rootrad}

\begin{figure}
  \begin{center}
    \includegraphics[width=\columnwidth]{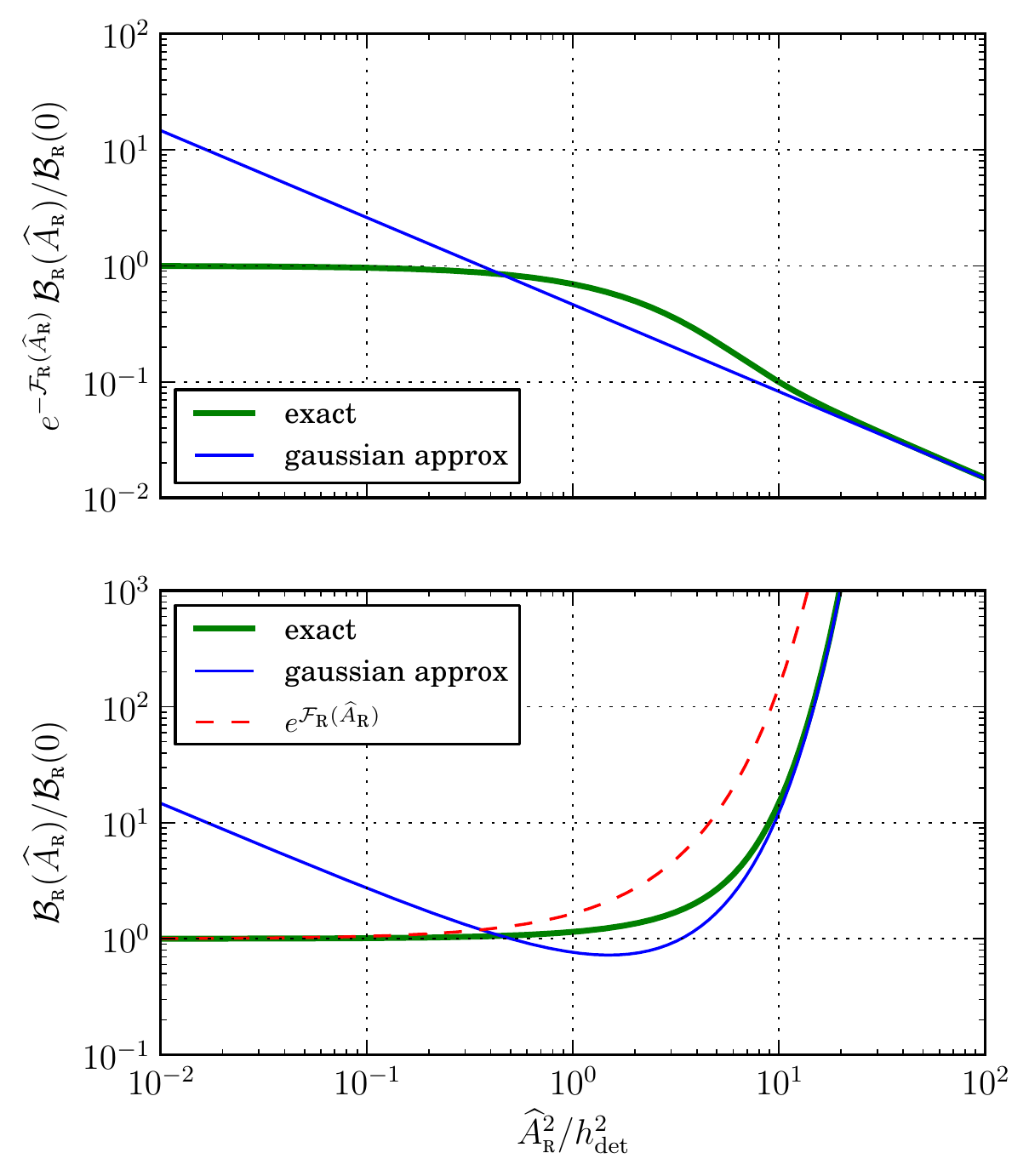}
  \end{center}
  \caption{Comparison of the exact form of
    $\frac{\BR(\MLEAR)}{\BR(0)}$ to the result of the
    approximate Gaussian integral, both with the factor of
    $e^{\F(\MLEAR)}=e^{\MLEAR/2\hdet^2}$ factored out and
    without.  Note that the value of the detection statistic matters,
    because the overall statistic is
    $\B(\detV{x})=\BR(\MLEAR)\BL(\MLEAL)$.}
  \label{f:hyperlim}
\end{figure}

We can compare the explicit result \eref{e:Bstatexplicit} to the
approximate result \eref{e:Bgauss} obtained in \sref{s:Bstat-rootrad}
by Gaussian integration in root-radius {\coord}s.  Applying the
explicit form \eref{e:explicitintegral} the Gaussian approximation
becomes
\begin{equation}
  \label{e:gaussaprox}
  \frac{\BR(\MLEAR)}{\BR(0)}
  \approx
  \frac{2\pi A \hdet^2}{\MLEAR^{3/2}}
  \,e^{\FsR(\MLEAR)}
  = \frac{2^{3/4}}{\Gamma(\frac{1}{4})}
  \left(\frac{\MLEAR}{\hdet}\right)^{\!-3/2}\! e^{\MLEAR^2/2\hdet^2}
  \ .
\end{equation}
We see that this agrees with the general result at large $\MLEAR$,
since
\begin{equation}
  {}_1F_1\left(\frac{1}{4},1,\frac{\MLEAR^2}{2\hdet^2}\right)
  \stackrel{\MLEAR\rightarrow\infty}{\longrightarrow}
  \frac{2^{3/4}}{\Gamma(\frac{1}{4})}
  \left(\frac{\MLEAR}{\hdet}\right)^{\!-3/2}\! e^{\MLEAR^2/2\hdet^2}
\end{equation}
by identity (13.5.1) of \cite{abramowitz64:_handb_mathem_funct}.
In \fref{f:hyperlim} we plot the exact form
\eref{e:hypersoln} of
$\frac{\BR(\MLEAR)}{\BR(0)}e^{-\F(\MLEAR)}$ as well as
the limiting form \eref{e:gaussaprox}.

\subsection{Range of Validity of CPF {\Coord} Gaussian
  Approximation}

\label{s:explicit-sigcoord}

Recall that in \sref{s:Bstat-sigcoord} we expand the combination
$\Lambda(\A;\detV{x})+\alpha(\A)$ about the maximum-likelihood point
$\A=\MLE{\A}$, where $\alpha(\A)$ is the logarithm of the measure of the
$\B$-statistic integral, given by \eref{e:alphadef}.  Subject to
the simplifying assumptions of this section, the log-likelihood ratio
becomes
\begin{equation}
  \begin{split}
    \Lambda(\A;\detV{x})
    & = \F(\detV{x})
    - \frac{1}{2\hdet^2}
    \bigl(
      [\Rx-\MLERx(\detV{x})]^2
      + [\Ry-\MLERy(\detV{x})]^2
      \\
      & \phantom{
        = \F(\detV{x})
        - \frac{1}{2\hdet^2}
        \bigl(
      }
      + [\Lx-\MLELx(\detV{x})]^2
      + [\Ly-\MLELy(\detV{x})]^2
    \bigr)
    \\
    &= \F(\detV{x})
    + \LambdaR(\Rx,\Ry;\detV{x})
    + \LambdaL(\Lx,\Ly;\detV{x})
  \end{split}
\end{equation}
and we can examine the behavior of
$\LambdaR(\Rx,\Ry;\detV{x})+\alphaR(\Rx,\Ry)$ and
$\LambdaL(\Lx,\Ly;\detV{x})+\alphaL(\Lx,\Ly)$ separately.  To
examine the integral for a particular data realization $\detV{x}$,
we can define rotated CPF {\coord}s
\begin{subequations}
  \label{e:Rp}
  \begin{align}
    \Rxp = \AR \cos(\tR-\MLEtR) &= \Rx\cos\MLEtR + \Ry\sin\MLEtR \\
    \Ryp = \AR \sin(\tR-\MLEtR) &= -\Rx\sin\MLEtR + \Ry\cos\MLEtR
  \end{align}
\end{subequations}
so that
\begin{multline}
  \LambdaR(\Rxp,\Ryp;\MLEAR)+\alphaR(\Rxp,\Ryp)
  \\
  = -\frac{1}{2\hdet^2}
  \left(
    [\Rxp - \MLEAR]^2 + [\Ryp]^2
  \right)
  -\frac{3}{4}\ln\left([\Rxp]^2+[\Ryp]^2\right)
  \ .
\end{multline}
We can find the stationary points explicitly, since
\begin{subequations}
  \begin{align}
    \frac{\partial(\LambdaR+\alphaR)}{\partial\Rxp}
    &= - \frac{\Rxp - \MLEAR}{\hdet^2}
    -\frac{3}{2}\frac{\Rxp}{(\Rxp)^2+(\Ryp)^2}
    \\
    \frac{\partial(\LambdaR+\alphaR)}{\partial\Ryp}
    &= - \frac{\Ryp}{\hdet^2}
    -\frac{3}{2}\frac{\Ryp}{(\Rxp)^2+(\Ryp)^2}
  \end{align}
\end{subequations}
we see that $\frac{\partial(\LambdaR+\alphaR)}{\partial\Ryp}=0$ when
$\Ryp=0$, which means that the stationary points occur when
\begin{equation}
  - \frac{\Rxp - \MLEAR}{\hdet^2}
  -\frac{3}{2\Rxp}
  = 0
\end{equation}
i.e., at the solutions of the quadratic equation
\begin{equation}
  2(\Rxp)^2 - 2\MLEAR(\Rxp) + 3 \hdet^2 = 0
\end{equation}
which are
\begin{equation}
  \Rxp
  = \frac{\MLEAR\pm\sqrt{\MLEAR^2-6\hdet^2}}{2}
\end{equation}
Since $\LambdaR(\Rxp,0;\MLEAR)+\alphaR(\Rxp,0)$ goes to
$+\infty$ as $\Rxp\rightarrow 0$ and $-\infty$ as
$\Rxp\rightarrow+\infty$, it is apparent that $\Rxp =
\frac{\MLEAR-\sqrt{\MLEAR^2-6\hdet^2}}{2}$
is a local minimum and $\Rxp = \Rxp_{\text{max}} =
\frac{\MLEAR+\sqrt{\MLEAR^2-6\hdet^2}}{2}$
is a local maximum.

\begin{figure}
  \begin{center}
    \includegraphics[width=\columnwidth]{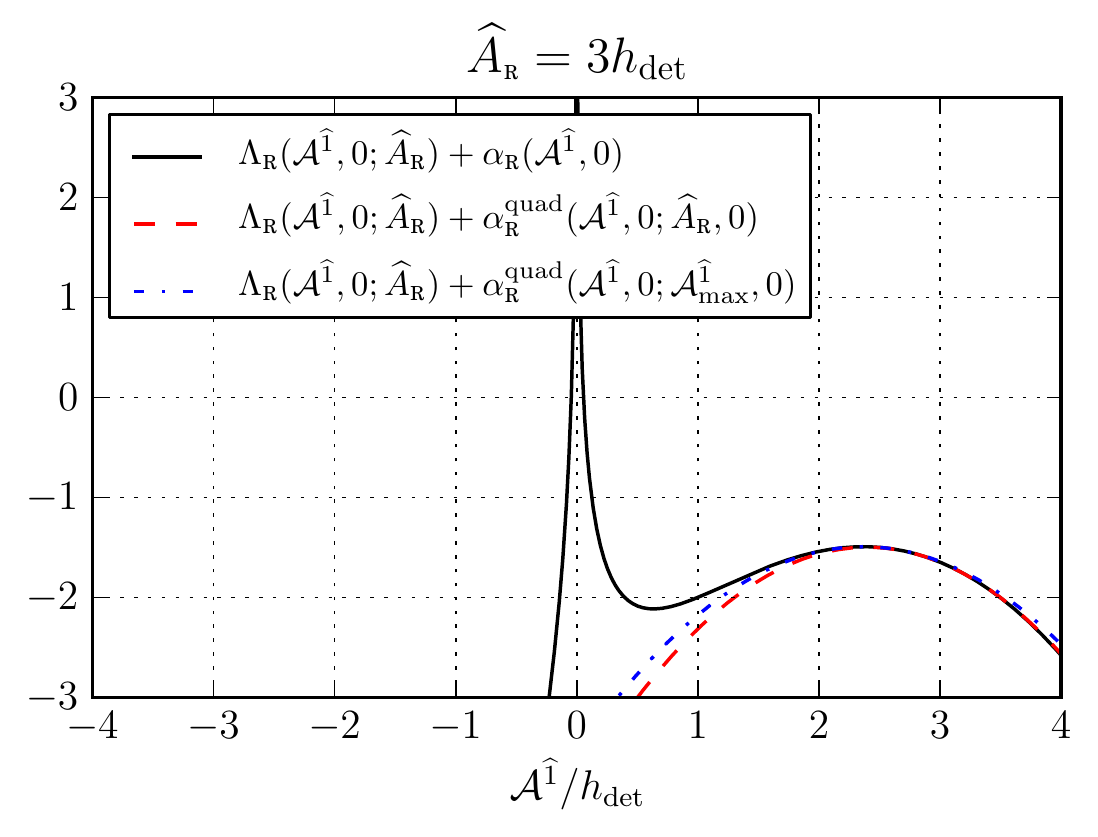}
  \end{center}
  \caption{Quadratic approximations to
    $\LambdaR(\Rxp,0;\MLEAR)+\alphaR(\Rxp,0)$ when $\MLEAR=3\hdet$, as
    a function of the rotated CPF {\coord} $\Rxp=\AR \cos(\tR-\MLEtR)$
    defined in \eref{e:Rp}, along the line $\Ryp=\AR
    \sin(\tR-\MLEtR)=0$.  We show two different quadratic
    approximations: one based on expanding $\alphaR(\Rxp,0)$ about the
    maximum likelihood point $\Rxp=\MLEAR=3\hdet$ and the other about
    $\Rxp_{\text{max}}=\frac{3+\sqrt{3}}{2}\approx 2.366\hdet$, at which
    $\LambdaR(\Rxp,0;\MLEAR)+\alphaR(\Rxp,0)$ has a local maximum.
    [Note that
    $\LambdaR(\Rxp,\Ryp;\MLEAR)+\alphaR^{\text{quad}}(\Rxp,\Ryp;\MLEAR,0)$
    has a local maximum at $(\Rxp,\Ryp)=(2.4\hdet,0)$.]}
  \label{f:quad_p_3}
\end{figure}

We also see that if $\MLEAR/\hdet<\sqrt{6}\approx 2.450$,
\emph{there is no local maximum}, only the singularity at the origin
of the $\{\Rxp,\Ryp\}$ plane.  Note that this condition is
actually more restrictive than the one corresponding to a saddle point
in the quadratic expansion at the maximum likelihood point.  That is
determined by the sign of
\begin{equation}
  \mc{N}_{\onedot\onedot}
  = -\frac{1}{\hdet^2} + \frac{3}{2\MLEAR^2}
\end{equation}
To follow the calculation of \sref{s:Bstat-sigcoord}, we define the
quadratic expansion of $\alphaR(\Rxp,\Ryp)$ about a point
$(\Rxp_0,\Ryp_0)$ as
\begin{multline}
  \alphaR^{\text{quad}}(\Rxp,\Ryp;\Rxp_0,\Ryp_0)
  \\
  = \alphaR(\Rxp_0,\Ryp_0)
  + \alphaR{}_{,\muhat}(\Rxp_0,\Ryp_0)[\A^{\muhat}-\A^{\muhat}_0]
  \\
  + \frac{1}{2}\alphaR{}_{,\muhat\nuhat}(\Rxp_0,\Ryp_0)
  [\A^{\muhat}-\A^{\muhat}_0][\A^{\nuhat}-\A^{\nuhat}_0]
\end{multline}
In particular
\begin{multline}
  \alphaR^{\text{quad}}(\Rxp,\Ryp;\Rxp_0,0)
  = -\frac{3}{2}\ln \Rxp_0
  \\
  - \frac{3}{2\Rxp_0}(\Rxp-\Rxp_0)
  + \frac{3}{4[\Rxp_0]^2}(\Rxp-\Rxp_0)^2
\end{multline}
If $\MLEAR/\hdet>\sqrt{6}\approx 2.450$, so that $\Rxp_{\text{max}} =
\frac{\MLEAR+\sqrt{\MLEAR^2-6\hdet^2}}{2}$ is a real number, the point
$(\Rxp,\Ryp)=(\Rxp_{\text{max}},0)$ is a local maximum of
$\Lambda(\Rxp,\Ryp)+\alpha(\Rxp,\Ryp)$, and the quadratic expression
\begin{equation}
  \Lambda(\Rxp,\Ryp)+\alpha^{\text{quad}}(\Rxp,\Ryp;\Rxp_{\text{max}},0)
\end{equation}
is an approximation to $\Lambda(\Rxp,\Ryp)+\alpha(\Rxp,\Ryp)$ near its
local maximum at $(\Rxp,\Ryp)=(\Rxp_{\text{max}},0)$.  This is the
situation illustrated in \fref{f:quad_p_3}, which plots
$\Lambda(\Rxp,0;\MLEAR)+\alpha(\Rxp,0)$ and its various quadratic
approximations when $\MLEAR=3\hdet$.

\begin{figure}
  \begin{center}
    \includegraphics[width=\columnwidth]{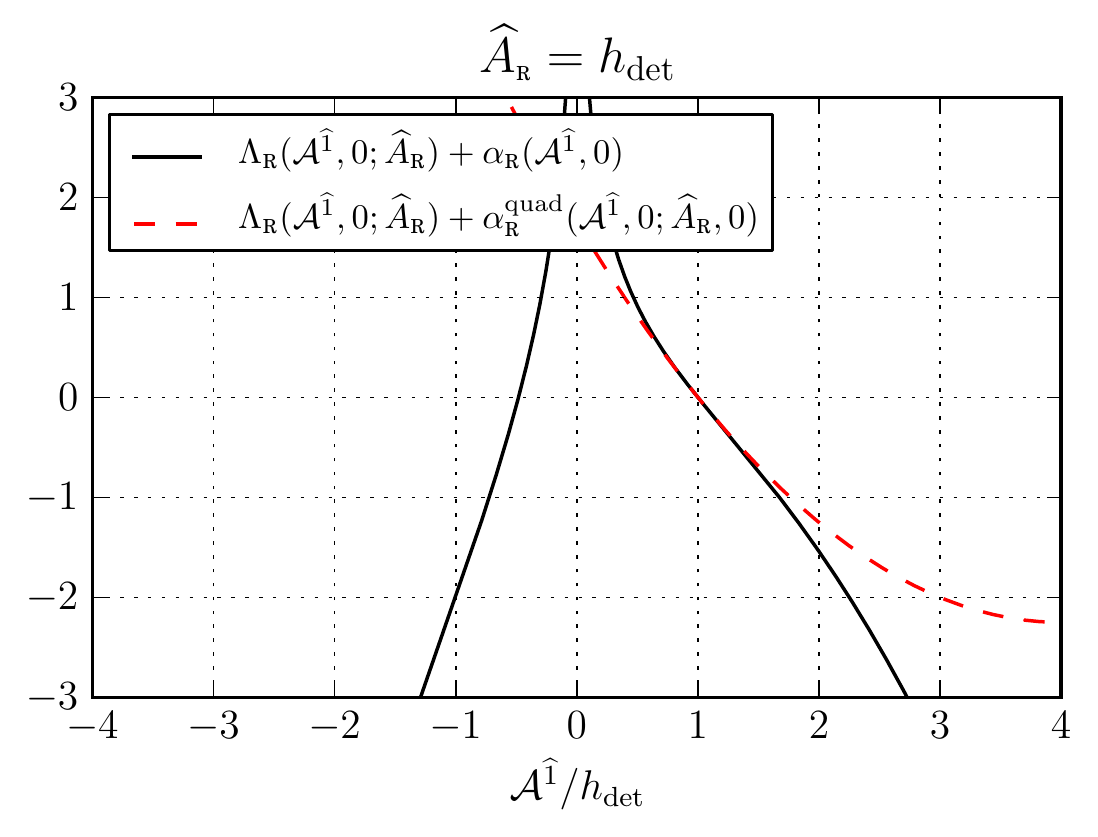}
  \end{center}
  \caption{Quadratic approximation to
    $\LambdaR(\Rxp,0;\MLEAR)+\alphaR(\Rxp,0)$ when
    $\MLEAR=\hdet$.  Here
    $\LambdaR(\Rxp,0;\MLEAR)+\alphaR(\Rxp,0)$ has no
    local maximum, only the singularity at $\Rxp=0$, and the
    stationary point of
    $\LambdaR(\Rxp,\Ryp;\MLEAR)+\alphaR^{\text{quad}}(\Rxp,\Ryp;\MLEAR,0)$,
    located at $(\Rxp,\Ryp)=(4\hdet,0)$, is a
    saddle point, since it curves upwards in the $\Rxp$ direction
    and downward in the $\Ryp$ direction.}
  \label{f:quad_p_1}
\end{figure}

We can always define a quadratic expansion about the maximum
likelihood point $(\Rxp,\Ryp)=(\MLEAR,0)$, namely,
\begin{multline}
  \Lambda(\Rxp,\Ryp)+\alpha^{\text{quad}}(\Rxp,\Ryp;\MLEAR,0)
  = \frac{\MLEAR^2}{2\hdet^2}-\frac{3}{2}\ln \MLEAR
  \\
  - \frac{3}{2\MLEAR}(\Rxp-\MLEAR)
  + \frac{1}{2}\left(\frac{3}{2[\MLEAR]^2}-\frac{1}{\hdet^2}\right)
  (\Rxp-\MLEAR)^2
  \\
\end{multline}
which will have a stationary point at $(\Rxp,\Ryp) =
\left(\MLEAR\left[\frac{6\hdet^2-2\MLEAR^2}{3\hdet^2-2\MLEAR^2}\right],0\right)$.
If $\MLEAR/\hdet<\sqrt{3/2}\approx 1.225$, this is a saddle point, as
illustrated in \fref{f:quad_p_1}, which plots
$\Lambda(\Rxp,0;\MLEAR)+\alpha^{\text{quad}}(\Rxp,0)$ and the
quadratic approximation
$\Lambda(\Rxp,0;\MLEAR)+\alpha^{\text{quad}}(\Rxp,0;\MLEAR,0)$ when
$\MLEAR=\hdet$.  Because the quadratic approximation curves upwards in
the $\Rxp$ direction, it cannot be used to calculate a Gaussian
integral.

\begin{figure}
  \begin{center}
    \includegraphics[width=\columnwidth]{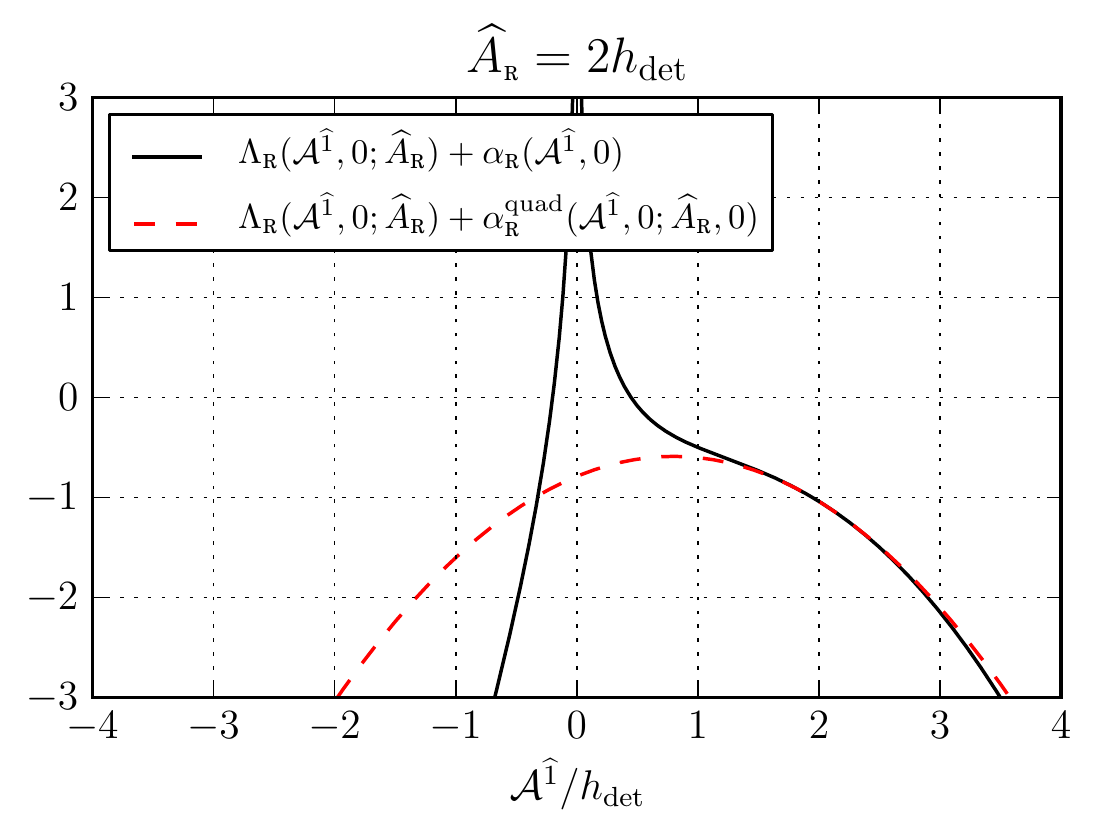}
  \end{center}
  \caption{Quadratic approximation to
    $\LambdaR(\Rxp,0;\MLEAR)+\alphaR(\Rxp,0)$ when
    $\MLEAR=2\hdet$.  Even though
    $\LambdaR(\Rxp,0;\MLEAR)+\alphaR(\Rxp,0)$ has no
    local maximum, the quadratic approximation
    $\LambdaR(\Rxp,\Ryp;\MLEAR)+\alphaR^{\text{quad}}(\Rxp;\MLEAR,0)$ has a
    local maximum at $(\Rxp,\Ryp)=(0.8\hdet,0)$.
    However, it is not a useful approximation to the
    original function for evaluating the integral, even though it is
    accurate close to the maximum likelihood point
    $\Rxp=2\hdet$.}
  \label{f:quad_p_2}
\end{figure}

\Fref{f:quad_p_2} shows an intermediate value $\MLEAR=2\hdet$, where
the quadratic approximation
$\LambdaR(\Rxp,\Ryp;\MLEAR)+\alphaR^{\text{quad}}(\Rxp,\Ryp;\MLEAR,0)$
has a local maximum, but $\LambdaR(\Rxp,\Ryp;\MLEAR)+\alphaR(\A)$ does
not.

\section{Conclusions}

We have demonstrated several new sets of {\coord}s on the amplitude
parameter space of continuous gravitational waves.  By taking linear
combinations \eref{e:PQ} of the usual Jaranowski-Kr\'{o}lak-Schutz
(JKS) {\coord}s, we obtain a set of variables, called CPF (circular
polarization factored) {\coord}s which are still {\coeff}s
in a linear representation \eref{e:tenswavedot} of the signal
waveform, but which are more closely connected to the physical
amplitude parameters of signal amplitude $h_0$, inclination and
polarization angles $\iota$ and $\psi$, and initial phase $\phi_0$.
In particular, these new {\coord}s divide naturally into two pairs of
Cartesian-like {\coord}s (one corresponding to right- and one to
left-circular polarization), and the polar {\coord}s
\eref{e:PQpolar} on these two subspaces, which we call CPF-polar
{\coord}s, are closely connected to the physical amplitude parameters:
the radial {\coord}s $\AR$ and $\AL$ are functions of $h_0$ and
$\cosi=\cos\iota$, while the angular {\coord}s $\tR$ and $\tL$ are
functions of $\psi$ and $\phi_0$, as shown in \eref{e:pqphys}.  We
also introduce so-called root-radius {\coord}s $\{\xR,\yR,\xL,\yL\}$,
derived from polar {\coord}s $\{\rR=\AR^{1/4},\tR,\rL=\AL^{1/4},\tL\}$,
which have the simplifying feature that the Jacobian of the
transformation between root-radius {\coord}s and the physical
{\coord}s $\{h_0,\cosi=\cos\iota,\psi,\phi_0\}$ is a constant.

We have presented several demonstrations of the utility of these new
{\coord}s.  They can be used in a simple derivation of the Jacobian
determinant \eref{e:Amu-phys} for the transformation between JKS and
physical {\coord}s (previously computed using computer algebra).  The
{\coord} singularities and ambiguities in physical parameters
associated with right or left circular polarization can be understood
as the origins of the two polar {\coord} systems,
\eref{e:rightcircpol} and \eref{e:leftcircpol}, respectively.
Finally, if we express in these {\coord}s the log-likelihood ratio
between models of Gaussian noise with and without a continuous
gravitational-wave signal, we can obtain results useful for the
calculation of the $\B$-statistic, which is the Bayes factor for a
comparison between the models.

Past work\cite{Prix09:_Bstat} has shown that if an unphysical prior is
used for the $\B$-statistic integral, an explicit Gaussian
integration in JKS {\coord}s (for which there is a straightforward
equivalent in CPF {\coord}s) shows that the $\B$ statistic is
equivalent to the $\F$ statistic.  If a more physically reasonable
prior is used, in particular one isotropic in the orientation angles
$\iota$ and $\psi$, the resulting Jacobian factor complicates the
evaluation of the integral.  We limited attention to the case where
the prior is uniform in the physical {\coord}s
$\{h_0,\cosi=\cos\iota,\psi,\phi_0\}$, and showed that the {\coord}
singularities in the resulting measure make even an approximate
Gaussian integration in CPF (or JKS) {\coord}s problematic.  We have
showed that an approximate Gaussian integration \emph{can} be
performed in root-radius {\coord}s, with the result that, up to an
irrelevant constant,
$\ln\B\approx\F-\frac{3}{2}\ln(\MLEAR\MLEAL)$, where
$\MLEAR$ and $\MLEAL$ are the maximum-likelihood estimates of the
CPF-polar radial {\coord}s $\AR=h_0\left(\frac{1+\cosi}{2}\right)^2$
and $\AL=h_0\left(\frac{1-\cosi}{2}\right)^2$.  This provides insights
into the $\B$ statistic in the regime where the signal is strong
and not too close to circular polarization.  Finally, we considered
the $\B$-statistic in the physical {\coord}s themselves and showed
that two of the four integrals could be performed exactly.

To gain more explicit insight into the behavior of the various
$\B$-statistic integrals, we considered a special case where the
amplitude parameter metric is diagonal, and showed that the simple
form of the log-likelihood ratio in this case allowed the integrals to
be performed analytically in CPF-polar {\coord}s, leading to an
explicit exact result \eref{e:Bstatexplicit} in terms of the
confluent hypergeometric function.  This could then be compared to the
approximate result from the Gaussian expansion in root-radius
{\coord}s to show the breakdown of the approximation for weak or
nearly-circularly-polarized signals.

\acknowledgments

The authors would like to thank Bruce Allen, Sanjeev Dhurandhar, Josh
Faber, Steve Fairhurst and Andy Lundgren for helpful discussions and
feedback.
JTW was supported by NSF grants PHY-0855494 and PHY-1207010.
RP and JLW were supported by the Max Planck Society.
CJC's work was carried out at the Jet Propulsion Laboratory,
California Institute of Technology, under contract to the National
Aeronautics and Space Administration; he gratefully acknowledges
support from NSF Grant PHY-106881.
This paper has been assigned LIGO Document Number \dcc, and
AEI-preprint number \aei.

\appendix

\section{Explicit form of Amplitude Parameter Metric}
\label{app:metric}

The explicit forms of the matrix elements \eref{e:Mmunu}
can be obtained explicitly by dividing the data from each detector $X$
into short stretches of data $[t_\iSFT,\,t_\iSFT + \Tsft)$ of length $\Tsft$ and Fourier
transformed (hence usually referred to as ``Short Fourier Transforms'',
or SFTs).
For a nearly monochromatic signal around frequency $f_0$, we can define
the usual (multi-detector) scalar product as
\begin{equation}
  \scalar{x}{y} \equiv \sumXiSFT \frac{4}{S^X_\iSFT(f_0)}
  \Real\int_{0}^{\infty} \cft{x}^{X*}_\iSFT(f) \, \cft{y}^{X}_\iSFT(f) \,df
\ ,
  \label{eq:def-scalar}
\end{equation}
where $S^X_\iSFT(f_0)$ is the one-sided noise power spectral density
around the frequency $f_0$ in detector $X$ during time stretch $\iSFT$,
and $\cft{x}_\iSFT^X, \cft{y}_\iSFT^X$ are the corresponding
Fourier-transforms of $x^X(t),y^X(t)$ restricted to the SFT
time-stretch $\iSFT$.

The explicit form of the scalar basis functions over each SFT
time stretch $t\in[t_\iSFT,\,t_\iSFT+\Tsft)$, defined according to
\eref{e:response} as $\cft{h}^X_{\mu,\iSFT}(f)\equiv
\cft{\tens{h}}{}^X_{\mu,\iSFT}(f):\tens{d}_\iSFT^X(f)$, can be given in the
time-domain as
\begin{subequations}
  \label{eq:hmu}
  \begin{align}
    h^X_{1,\iSFT}(t) &= a^X_\iSFT(f_0)\,\cos\phi^X(t)
\ , \\
    h^X_{2,\iSFT}(t) &= b^X_\iSFT(f_0)\,\cos\phi^X(t)
\ , \\
    h^X_{3,\iSFT}(t) &= a^X_\iSFT(f_0)\,\sin\phi^X(t)
\ , \\
    h^X_{4,\iSFT}(t) &= b^X_\iSFT(f_0)\,\sin\phi^X(t)
\ ,
  \end{align}
\end{subequations}
where we defined the shorthand $\phi^X(t)\equiv\phi(\tssb^X(t))$, and
where the frequency-dependent complex AM {\coeff}s for time stretch
$\iSFT$ are (see \eref{eq:def-a-b}) $a^X_\iSFT(f)=\epsp :
\tens{d}_\iSFT(f)$ and $b^X_\iSFT(f)=\epsc : \tens{d}_\iSFT(f)$,
respectively. These reduce to the real-valued constants $a^X_\iSFT$
and $b^X_\iSFT$ in the long-wavelength limit.
Using this together with the definition of the amplitude-parameter metric
\eref{eq:Mmunu} and \eref{e:Mmunu} we find
\begin{subequations}
  \begin{align}
    A &= \sumXiSFT\frac{\Tsft}{S^X_\iSFT(f_0)}\abs{a^X_\iSFT(f_0)}^2
\ ,
    \\
    B &= \sumXiSFT\frac{\Tsft}{S^X_\iSFT(f_0)}\abs{b^X_\iSFT(f_0)}^2
\ ,
    \\
    C &= \sumXiSFT \frac{\Tsft}{S^X_\iSFT(f_0)}\Real\left[a^{X*}_\iSFT(f_0)\,b^X_\iSFT(f_0)\right]
\ ,
    \\
    E &= \sumXiSFT \frac{\Tsft}{S^X_\iSFT(f_0)}\Imag\left[a^{X*}_\iSFT(f_0)\,b^X_\iSFT(f_0)\right]
\ ,
  \end{align}
\end{subequations}

The explicit forms of the matrix elements $\M_{\mudot\nudot} =
\scalar{h_{\mudot}}{h_{\nudot}}$ can be
obtained by using either \eref{e:dottenswf} or
\eref{e:circtenswf}, namely
\begin{subequations}
  \begin{align}
    \begin{split}
      I &= A+B+2E
      = \sumXiSFT \frac{\Tsft}{S^X_\iSFT(f_0)}
      \abs{a^X_{\Left,\iSFT}(f_0)}^2
\ ,
    \end{split}
    \\
    \begin{split}
      J &= A+B-2E
      = \sumXiSFT \frac{\Tsft}{S^X_\iSFT(f_0)}
      \abs{a^X_{\Right,\iSFT}(f_0)}^2
\ ,
    \end{split}
    \\
    \begin{split}
      K &= 2C
      = \sumXiSFT \frac{\Tsft}{S^X_\iSFT(f_0)}
      \Real\left[a^{X*}_{\Left,\iSFT}(f_0)\,a^X_{\Right,\iSFT}(f_0)\right]
\ ,
    \end{split}
    \\
    \begin{split}
      L &= A-B
      = \sumXiSFT \frac{\Tsft}{S^X_\iSFT(f_0)}
      \Imag\left[a^{X*}_{\Left,\iSFT}(f_0)a^X_{\Right,\iSFT}(f_0)\right],
    \end{split}
  \end{align}
\end{subequations}
where
\begin{subequations}
  \begin{align}
    a^X_{\Right,\iSFT}(f_0)=\epsr:\tens{d}^X_{\iSFT}(f_0)
    &= a^X_\iSFT(f_0) + ib^X_\iSFT(f_0)
\ ,
    \\
    a^X_{\Left,\iSFT}(f_0)=\epsl:\tens{d}^X_{\iSFT}(f_0)
    &= a^X_\iSFT(f_0) - ib^X_\iSFT(f_0)
\ .
  \end{align}
\end{subequations}
Note that only in the long-wavelength limit we have
$a^{X*}_{\Right,\iSFT}=a^X_{\Left,\iSFT}$ and therefore also $I=J$.

Note that the scalar CPF waveforms of
\eref{e:dottenswf},\eref{e:circtenswf} can be obtained as
\begin{subequations}
  \begin{align}
    h^X_{\onedot,\iSFT}(t)
    &= \frac{1}{2}\left( a^X_{\Right,\iSFT}(f_0)\,e^{-i\phi^X(t)} + a^X_{\Left,\iSFT}(f_0)\,e^{i\phi^X(t)}\right)
\ ,\\
    h^X_{\twodot,\iSFT}(t)
    &= \frac{1}{2i}\left( a^X_{\Right,\iSFT}(f_0)\,e^{-i\phi^X(t)} - a^X_{\Left,\iSFT}(f_0)\,e^{i\phi^X(t)}\right)
\ ,\\
    h^X_{\tredot,\iSFT}(t)
    &= \frac{1}{2} \left( a^X_{\Left,\iSFT}(f_0)\,e^{-i\phi^X(t)} + a^X_{\Right,\iSFT}(f_0)\,e^{i\phi^X(t)}\right)
\ ,\\
    h^X_{\fordot,\iSFT}(t)
    &= \frac{1}{2i} \left(a^X_{\Left,\iSFT}(f_0)\,e^{-i\phi^X(t)} + a^X_{\Right,\iSFT}(f_0)\,e^{i\phi^X(t)}\right)
\ ,
  \end{align}
\end{subequations}
which does not share the simple form of \eref{e:circtenswf}, as
the detector tensor $\tens{d}_\iSFT^X(f)$ is generally complex.
However, the detector response in the time-domain is real-valued, and
therefore
$\tens{d}^{X*}_\iSFT(f) = \tens{d}^X_\iSFT(-f)$, and also $a^{X*}_{\Right,\iSFT}(f) = a^X_{\Left,\iSFT}(-f)$.

\section{Hyperbolic {\coord}s}

\label{app:hyp}

Here we present an additional {\coord} systems which has the
simplifying feature that the Jacobian to transfer between it and the
physical {\coord}s $\{h_0,\cosi,\psi,\phi_0\}$ is a constant.

If we consider the amplitudes
\begin{equation}
  \Ap = \frac{h_0}{2} (1+\cosi^2)
  \qquad\hbox{and}\qquad
  \Ac = h_0 \cosi
\end{equation}
and note that
\begin{equation}
  \Ap^2 - \Ac^2 = \frac{h_0^2}{4}(1-\cosi^2)^2 > 0
\end{equation}
it seems natural to define
\begin{subequations}
  \begin{align}
    \HA &= \sqrt{\Ap^2 - \Ac^2} = \frac{h_0}{2}(1-\cosi^2)
    \\
    \e &= \frac{1}{2} \ln \frac{\Ap+\Ac}{\Ap-\Ac} = \ln\frac{1+\cosi}{1-\cosi}
  \end{align}
\end{subequations}
so that
\begin{equation}
  \Ap = \HA\coshe
  \qquad\hbox{and}\qquad
  \Ac = \HA\sinhe
  \ .
\end{equation}
We can invert the {\coord} transformations to show
\begin{equation}
  h_0 = \HA (1+\cosh\e)
  \qquad\hbox{and}\qquad
  \cosi = \tanh \frac{\e}{2}
  \ .
\end{equation}

The Jacobians between $\{\Ap,\Ac\}$ and the two {\coord} systems
$\{h_0,\cosi\}$ and $\{\HA,\e\}$ give
\begin{equation}
  \HA\,d\HA\,d\e = d\Ap\, d\Ac = \frac{h_0}{2}(1-\cosi^2)\,dh_0\,d\cosi
\end{equation}
from which we see
\begin{equation}
  d\HA\,d\e = dh_0\,d\cosi
\end{equation}

Note that in these hyperbolic {\coord}s, circular polarization is not
represented by finite {\coord} values.  As $\cosi\rightarrow 1$ at
finite $h_0$, $\HA\rightarrow 0$ and $\eta\rightarrow\infty$ so that
\begin{equation}
  \Ap\rightarrow \HA \frac{e^\e}{2} \qquad \Ac\rightarrow \HA \frac{e^\e}{2}
  \ .
\end{equation}
As $\cosi\rightarrow -1$ at
finite $h_0$, $\HA\rightarrow 0$ and $\eta\rightarrow -\infty$ so that
\begin{equation}
  \Ap\rightarrow \HA \frac{e^{-\e}}{2}
  \qquad \Ac\rightarrow -\HA \frac{e^{-\e}}{2}
  \ .
\end{equation}

\bibliography{biblio}

\end{document}